\documentclass[sn-nature]{sn-jnl}
\usepackage{times}

\usepackage{subcaption} 
\usepackage{graphicx}
\usepackage[strict]{changepage}
\usepackage{multirow} 
\usepackage{multicol} 
\usepackage{array} 
\usepackage{tabularx} 
\usepackage{verbatim} 

\usepackage[dvipsnames]{xcolor}
\usepackage{makecell}
\usepackage{amsmath,amstext,amssymb}
\usepackage{amsthm}%
\usepackage{booktabs}
\usepackage{threeparttable} 
\usepackage{csquotes}
\usepackage{todonotes}
\usepackage[title]{appendix}
\usepackage{longtable}
\usepackage{textcomp}%
\usepackage{manyfoot}%
\usepackage{xr-hyper, hyperref}

\usepackage{geometry}

\newtheorem{hyp}{Hypothesis}
\def\sym#1{\ifmmode^{#1}\else\(^{#1}\)\fi}

\usepackage[normalem]{ulem}

\newcommand{\weic}[1]{{\small{\color{Mahogany}{[wei: #1]}}}}

\usepackage{xcolor}
\definecolor{BrightBlue}{RGB}{0, 191, 255} 

\usepackage[most]{tcolorbox}
\usepackage{float}

\date{\today}

\begin{document}



\title{Using Artificial Intelligence to Unlock Crowdfunding Success for Small Businesses} 



\author[1]{\fnm{Teng} \sur{Ye}$^{\dagger}$ }

\author[2]{\fnm{Jingnan} \sur{Zheng} $^{\dagger}$}

\author[3]{\fnm{Junhui} \sur{Jin}}

\author[4]{\fnm{Jingyi} \sur{Qiu}}

\author[5]{\fnm{Wei} \sur{Ai}}

\author*[4]{\fnm{Qiaozhu} \sur{Mei}}

\affil[1]{\orgdiv{Carlson School of Management}, \orgname{University of Minnesota}}

\affil[2]{\orgdiv{School of Computing}, \orgname{National University of Singapore}}

\affil[3]{\orgdiv{Electrical Engineering and Computer Science}, \orgname{University of Michigan}}

\affil*[4]{\orgdiv{School of Information}, \orgname{University of Michigan}}

\affil[5]{\orgdiv{College of Information Studies}, \orgname{University of Maryland}}













\baselineskip24pt



\abstract{
While small businesses are increasingly turning to online crowdfunding platforms for essential funding, over 40\% of these campaigns may fail to raise any money, especially those from low socio-economic areas. We utilize the latest advancements in AI technology to identify crucial factors that influence the success of crowdfunding campaigns and to improve their fundraising outcomes by strategically optimizing these factors.  Our best-performing machine learning model accurately predicts the fundraising outcomes of 81.0\% of campaigns, primarily based on their textual descriptions. Interpreting the machine learning model allows us to provide actionable suggestions on improving the textual description before launching a campaign. We demonstrate that by augmenting just three aspects of the narrative using a large language model, a campaign becomes more preferable to 83\% human evaluators, and its likelihood of securing financial support increases by 11.9\%. Our research uncovers the effective strategies for crafting descriptions for small business fundraising campaigns and opens up a new realm in integrating large language models into crowdfunding methodologies.}

\keywords{Crowdfunding, Small Business, AI, Large Language Models}


\maketitle

\newpage
\section{Introduction}
Small businesses provide a critical foundation for the US economy \cite{forbes_2022}. According to the U.S. Small Business Association (SBA), small businesses consist of 99.9\% of American businesses, which together employed 61.7 million Americans as of 2023 \cite{sba_2023}. Unlike larger corporations, small businesses are more financially vulnerable to policy changes and economic recessions, and they have access to fewer financial support resources \cite{welsh1981small,sahin2011small}. This challenge is particularly exaggerated during economic shocks, such as the 2008 financial crisis, the COVID-19 lockdown, and natural disaster occurrences. In these situations, small businesses are found more financially fragile and more likely to face permanent closure than large firms (e.g., \citenum{duygan2015financing,tierney1997business, bartik2020impact,fairlie2023were}). While governments strive to provide funding assistance, numerous small businesses encounter challenges in accessing these programs due to bureaucratic complexities and hurdles in proving their eligibility \cite{bartik2020impact}.

Small businesses are increasingly resorting to crowdfunding platforms to secure financial support. Projections estimate that the annual volume of such funding will reach \$6.8 billion by 2031 \cite{forbes_2023}. Nevertheless, turning to crowdfunding presents its own set of challenges, as many practitioners have limited experience and knowledge in organizing successful crowdfunding campaigns. About 40\% of these campaigns did not receive any funding \cite{igra2021crowdfunding}. To address this challenge, prior studies have investigated various campaign strategies to facilitate crowd fundraising, including the use of matching donations \cite{chen2005online,karlan2007does} and offering thank-you gifts \cite{falk2007gift, alpizar2008anonymity, chao2017demotivating}. These mechanisms, however, typically tested one at a time, 
may face limited real-world applicability due to financial and logistical constraints. In contrast, business owners usually have more control and flexibility over the narratives of their campaigns, allowing them to incorporate practical linguistic enhancements into their crafted messages. This includes making their narratives more emotionally compelling \cite{rhue2018emotional} and socially oriented \cite{zhang2021contributes,du2021predicting}, focusing more on the benefits to donors or recipients \cite{list2021experimental}, and using less common words to convey their stories \cite{markowitz2021predictive}. 


However, in the absence of accurate methods for text comprehension, previous research on linguistic factors has been restricted to lexical and syntactic analysis. This limitation prevents a complete understanding of the extensive scope, depth, and complexities of linguistic expression within narratives. Such analyses typically fail to provide sufficient insights into the strategic presentation of fundraising campaigns, including critical aspects such as budget plans and business history. Consequently, these elements are often examined in a piecemeal fashion and/or experimented on a case-by-case basis.



The recent advancements in large language models have unveiled new opportunities, thanks to their compelling capabilities in text comprehension and generation. As shown in recent studies, the ChatGPT operated by OpenAI outperforms human crowdworkers in a number of text annotation tasks \cite{gilardi2023chatgpt} and boosts worker productivity in certain writing tasks \cite{noy2023experimental}. 

Drawing on these findings and theoretical insights, we identify and quantify over a hundred aspects in the descriptions of small business fundraising campaigns with the help of large language models and lexical tools. 
By employing these identified cues as features, we utilize machine learning models to forecast the outcomes of these campaigns. Our most effective model achieves an accuracy of 81.0\% in predicting fundraising results. By evaluating the importance of various features, we have compiled a detailed list of factors that influence the success of crowdfunding campaigns. This analysis provides actionable guidance on how to introduce the business, explain the urgency of the request, and express appreciation effectively in the campaign introduction. 

Our study demonstrates that ChatGPT can be effectively instructed to revise campaign introductions by incorporating these insights. A simulation analysis suggests that a campaign could increase its likelihood of receiving financial support by an average of 11.9\%, after integrating just three of these key factors. This benefit is significantly higher on originally unfunded campaigns.  Our findings are further supported by an online experiment, where a revised campaign introduction is preferable by 83\% human evaluators. 

Our research marks a pioneering step in the use of generative AI for crowdfunding. We've developed a pipeline that employs large language models to extract key factors from campaign narratives and to implement these factors. This approach is accessible and feasible for small businesses, even those without extensive knowledge of campaign organization or natural language processing. Our findings offer practical solutions for optimizing fundraising campaigns, even for those lacking the resources to deploy an AI-augmented approach. They can significantly boost the success rate of their campaign by simply revising a few key aspects in its introduction before launch. Our research contributes to enhancing the overall success of small business crowdfunding campaigns as well as increasing the equity in this context.



\section{Methods}
\begin{figure}[ht]
\centering
\includegraphics[width=\textwidth]{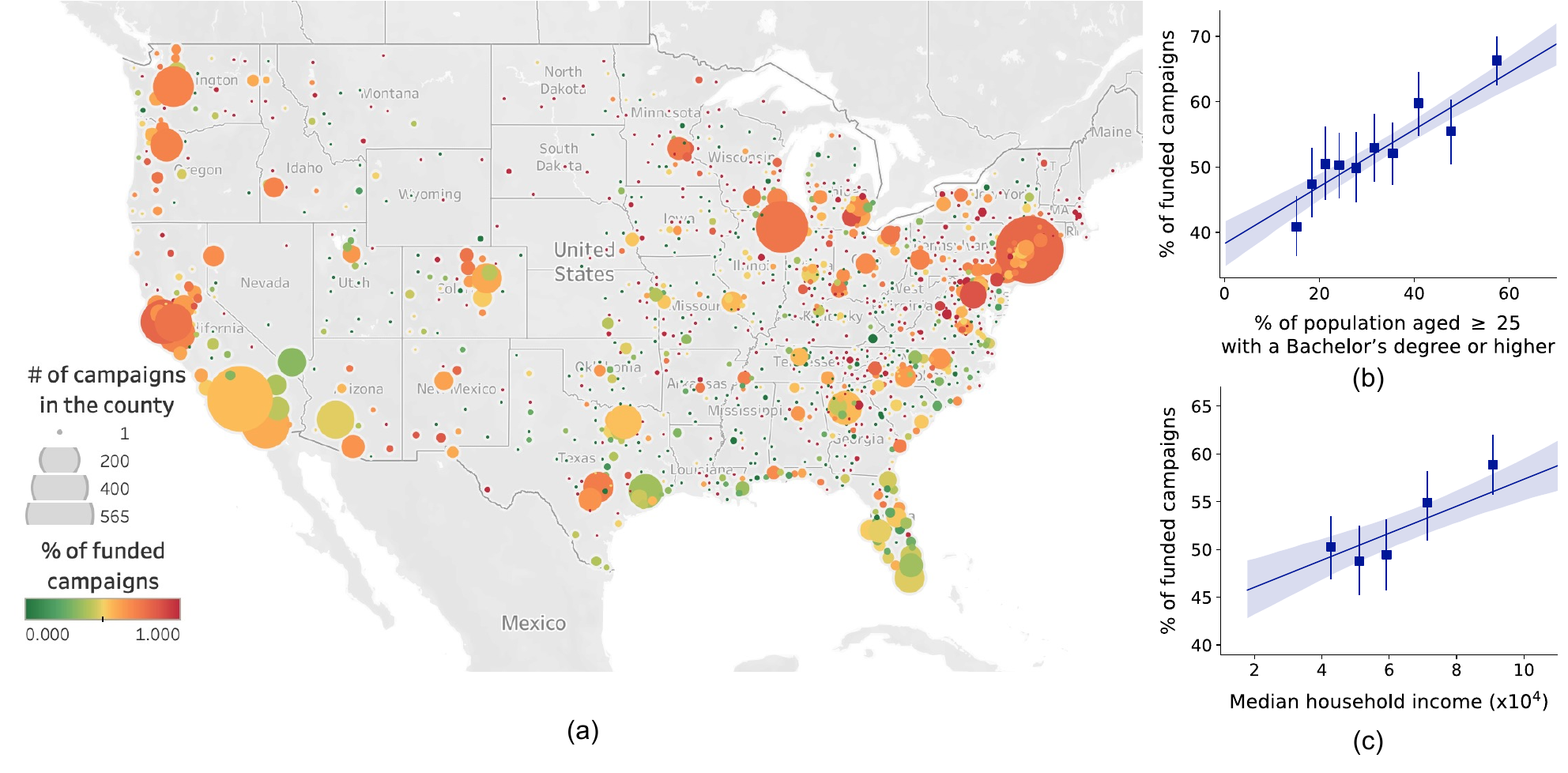}
\caption{Overview of the Distribution and Funded Rates of Small Business Crowdfunding Campaigns. The left panel (a) displays the geographical distribution at the county level with the size of circle representing the total number of campaigns and the color describing the average percentage of funded campaigns. The right panels (b) and (c) showcase the relationship between local socio-economic status and the likelihood of receiving funds at the city level. (b) indicates that the percentage of residents' with higher education in a city is positively related to the funded probability. Similarly, (c) suggests a positive correlation between the city median household income and the percentage of campaigns receiving funds.}
\label{fig:overview}
\end{figure}

We collect 11,274 fundraising campaigns that fall into the small-business-relief collection on GoFundMe, the largest crowdfunding platform in the world \cite{usnews_2023}. These campaigns encompass 50 States in the U.S. and the period between January $22^{nd}$, 2020 and December $31^{st}$, 2020, which corresponds to the initial outbreak and lockdowns of the COVID-19 pandemic, the most recent external shock that has broad impact on small businesses (\ref{si-sec:data_collection}). Figure \ref{fig:overview} presents an overview of the geographical distribution of these campaigns and the ratio of funded campaigns per county. Our results indicate geographical disparities in crowdfunding success, with campaigns originating from regions of lower socio-economic status (in terms of education level and median household income) facing lower percentage of receiving funding. 

On GoFundMe, fundraisers seek donations from the general public by posting a campaign that specifies the fundraising goal amount, relevant photo(s), and a short textual description. We combine behavioral theories and computational tools to identify the key factors that contribute to fundraising success. These factors are derived from four sources: the textual description of the campaign, campaign configuration (such as the targeted fundraising goal amount), demographics in the business's local area (from the American Community Survey), and other contextual factors (such as the severity of the pandemic shock). The detailed feature extraction process is described in the Supplementary Information (\ref{SI:features}).

We primarily focus on the factors derived from the text description, as these are easily adjustable by fundraisers before launching their campaign and the text description serves as the main channel through which potential donors gather crucial information about the business and the campaign \cite{zhou2018project}. Such information includes %
the nature of the small business's services, the mutual benefits for both the business and its donors \cite{list2021experimental}, and the reasons why the small business is requesting the specified financial support. Drawing on insights from prior literature in crowdfunding, small business, and natural language processing (e.g., \citenum{list2021experimental, frey2004social,markowitz2021predictive, rhue2018emotional, tan2016winning,brysbaert2014concreteness, warriner2013norms,wang2015making}), we employ both large language models (LLMs) and lexicon-based methods (LBMs) to extract factors from campaign descriptions. LLMs assist in capturing semantic and pragmatic factors such as social comparison with peer businesses, the plan for funding usage, and the urgency of fundraising. LBMs help in depicting the writing style and sentiment of the description. A total of 11 factors are extracted using large language models, and 105 features are identified through lexicon-based methods. These factors are then utilized as features in a predictive analysis of the fundraising outcomes (See \ref{sec:text} for a complete list of textual features.)

Our data, in line with prior literature \cite{igra2021crowdfunding}, shows that more than 40\% of campaigns received no funding at all.  We therefore predict the fundraising outcome as a binary value, indicating whether or not the small business secured any financial support from its crowdfunding campaign on GoFundMe.  \footnote{We collected campaigns created from the beginning of the COVID-19 pandemic to December $31^{st}$, 2020 and followed their daily status updates until Feb $5^{th}$, 2021). This should give us a reliable estimate of the fundraising outcome given that 93.98\% GoFundMe small-business-relief campaigns that received financial support had their first donation within a month.} 

We employ the Light Gradient-Boosting Machine (LightGBM) \cite{ke2017lightgbm}, a non-linear tree-based machine learning model known for its high explanability and computational efficiency, to develop a predictor of this binary outcome. We follow the standard practices of machine learning to divide the data into training, validation, and test sets (Table \ref{tab:data_split}) according to the rule of time-dependent knowledge restriction and rigorously tune the hyperparameters (more details in \ref{si-sec:prediction_model}).

The machine learning model measures the importance of individual features in predicting the fundraising outcome (\ref{si-sec:prediction_model}). Through the OpenAI API \cite{OpenAI}, we prompt ChatGPT-4 to rewrite 500 campaign descriptions by adjusting particular aspects in the narrative that correspond to a subset of the important features identified by the best-performing prediction model (\ref{si:simulation_rewrite}). 

The AI-revised campaign descriptions are evaluated via two complementary analyses: first, an offline simulation analysis that estimates the counterfactual likelihood of a GPT-augmented campaign receiving funding (\ref{si:simulation}); and second, an online randomized experiment that directly contrasts the campaign descriptions before and after AI augmentation (\ref{si:experiment}).
\section{Results}
\label{sec:results}

\subsection{Prediction Results}

Our best performing LightGBM model achieves 81\% accuracy in predicting the fundraising outcomes of out-of-sample small business campaigns, marking an 36.59\% improvement over a baseline uniform guess. More details are shown in Table \ref{tab:lgb}. 

\begin{figure}[ht]
\centering
\includegraphics[width=1\textwidth]{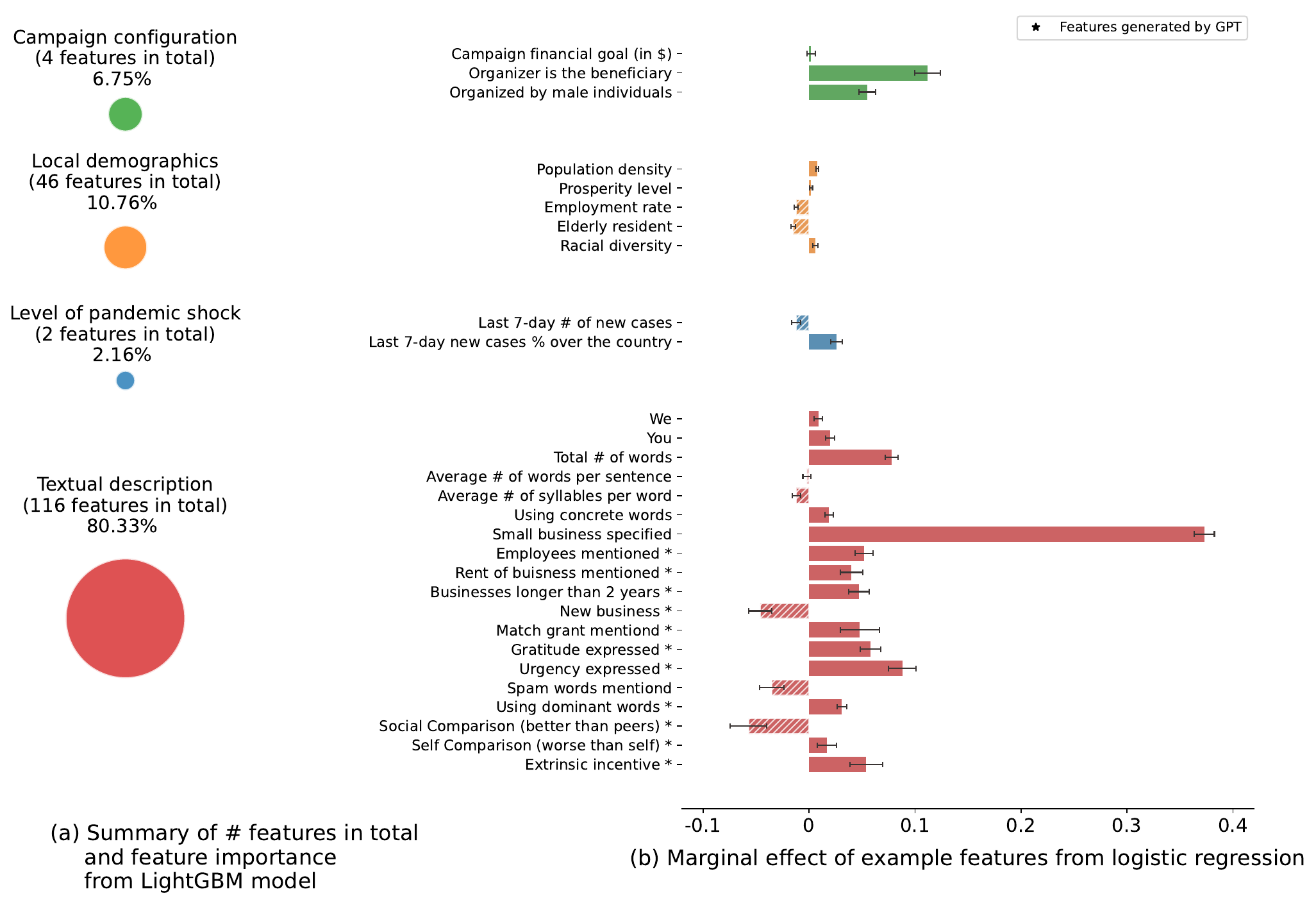}
\caption{Summary and Examples of Feature Importance. (a) displays the proportion of importance for each group of features in the LightGBM prediction model. (b) showcases the average marginal effect with \textit{delta-method} (SE in error-bars), where features in the same group are depicted with the same color. This figure indicates significant influence of textual features: they account for 80.33\% of the decisive power among all feature groups, including local demographics, pandemic shock, and GoFundMe campaign configurations. }
\label{fig:feature_imp}
\end{figure}

By interpreting the feature importance of the best-performing model, which quantifies the increase of information gain achieved by each feature \footnote{LightGBM trains an ensemble of decision trees, and information gain of a feature meausures the loss reduction caused by using the feature to split each decision tree (i.e., reduction of  prediction error by splitting a node into two leaves using the feature), aggregated across all trees.}, we are able to identify the most predictive features for the fundraising outcome. As shown in Figure \ref{fig:feature_imp} (a), we find that textual features extracted from the description collectively account for 80.33\% of the improvement in predictive accuracy towards fundraising success, followed by the demographics of the local area (10.76\%),  campaign configuration (6.75\%), and the severity of the pandemic shock to the local area (2.16\%). The importance of the textual features is validated by an ablation study that compares the performance of the prediction model with or without the group of features (\ref{tab:lgb_ablation}). The importance of individual features is more skewed. One single feature, a confirmation of the small business identity of the fundraising entity in the campaign description, alone contributes to 36\% of the improvement in predictive accuracy. 

To discern the specific relationships between individual textual features and the fundraising outcome, we examine the features with high importance in the best-performing LightGBM model. We then train a Logistic Regression model to understand the contribution and influence of these features on the campaign success (\ref{si-sec:logit}). The main results are summarized in Figure~\ref{fig:feature_imp}(b) and details are reported in Table \ref{tab:regression}. Table \ref{tab:feature_literature_finding} highlights the related literature and our findings regarding the textual features.

The result of logistic regression corroborates several findings from the existing literature. For example, longer descriptions are associated with a higher likelihood of donation to campaigns ($p<.01$). The finding is consistent with previous research on Kickstarter, a crowdfunding platform where donors typically get a copy of product in exchange for their financial support \cite{duan2020entrepreneurs}. In line with previous studies on the effects of personal pronouns \cite{zhu2022proximal}, our results (Table \ref{tab:regression} Model (3)) show positive correlations between the success of fundraising and the presence (percentage) of the word \textit{``We''} ($\beta=0.01$, i.e., 1\% of increase, $p<.01$) and the word \textit{``You''} ($\beta=0.02$, i.e., 2\% of increase, $p<.01$) in the campaign description.

Word usage has other implications for fundraising success. Our results show that campaign descriptions that use words with more syllables on average are less likely to receive funding than those using words with fewer syllables ($\beta=-0.01$, $p<.01$). This finding contrasts with previous research on Kickstarter \cite{duan2020entrepreneurs}. This discrepancy may stem from the fact that Kickstarter predominantly caters to creative projects, where descriptions using elaborate and intricate language might be more appealing to the audience. On the other hand, small-business campaigns on GoFundMe generally seek funding for weathering difficult economic circumstances. In this context, simple and straightforward language may be more effective in conveying their immediate needs and the urgency of their situation. In a similar vein, descriptions utilizing more concrete words are more likely to receive funds ($\beta=0.02$, $p<.01$). 

Our findings reveal that words that are highly suggestive of spam, detected by the package referenced in \cite{wang2015making}, reduce the likelihood of a campaign to secure financial support ($\beta=-0.04$, $p<.01$). This underscores the critical need for legitimate small-business fundraisers to clearly distinguish their campaigns from fraudulent ones.

Beyond these lexical characteristics, the semantic and pragmatic features identified by large language models unveil new insights previously unreported in crowdfunding literature, as detailed in Model (3) in Table \ref{tab:regression}. These features offer a deeper understanding of the nuanced and context-dependent aspects of language that influence crowdfunding success. We find that small businesses who identified themselves as in operation for more than 2 years or highlighted their history are more likely to receive funding ($\beta=0.05$, $p<.01$). 
Donors are less likely to invest in new businesses ($\beta=-0.05$, $p<.01$) when it is specified in the description. Campaigns that specifically mentioned their purposes of raising funds as covering rent ($\beta=0.04$, $p<.01$) or supporting employee ($\beta=0.05$, $p<.01$), are more likely to raise funds. Corroborating prior studies \cite{falk2007gift}, when provided extra incentive, such as thank you gifts and store gift card, donors are more willing to support the campaign.

Surprisingly, descriptions that imply self-comparison or social comparison show opposite correlations with the campaign outcome. Businesses that communicated a worsened situation compared to their past (self comparison) showed a marginally significant positive correlation with fundraising success ($\beta=0.02$, $p<.10$). However, those who portrayed themselves as superior to their peers in terms of product or service quality (social comparison) are less likely to receive funding support ($\beta=-0.06$, $p<.01$). This intriguing finding extends literature on social comparison theory in crowdfunding. While previous research emphasizes the positive effect of presenting the donation behaviors of a donor's peers on individual donor contributions \cite{frey2004social,shang2009field}, our study focuses on the fundraisers' social information release, instead of the donors, shedding light upon the potential adverse impact of social comparison.  

Notably, some of our findings have practical implications for a wide array of (if not all) campaigns. For example, our result suggests that campaigns that clearly expressed gratitude to potential donors ($\beta=0.06$, $p<.01$) and those who explained the urgency of their funding needs ($\beta=0.09$, $p<.01$) are more likely to receive donation. Yet, only 62.9\% and 79.6\% of the campaigns actually implemented these strategies respectively. Additionally, campaigns that acknowledged the availability of match grants in descriptions were 4.8\% ($p<.01$) more likely to secure financial support. Despite this clear advantage, only 9.3\% of small businesses mentioned match grants in their campaigns. This is notable considering the existence of a platform-wide matching mechanism on GoFundMe that offers an additional \$500 for every small business who can raise at least \$500. These strategies can be readily implemented in a campaign description either manually or through a language language model. 

\subsection{Counterfactual Forecasting}
\label{sec:simulation}

We prompt ChatGPT-4 to revise existing campaign descriptions, aligning them with three optimal strategies we've identified above: (1) express gratitude to donors; (2) explain the urgency of funding needs; (3) acknowledge the availability of matching grants. These modifications alter the language of a campaign while preserving its fundamental characteristics. The alterations made to the campaign description are expected to modify its textual features. This, in turn, could potentially result in a counterfactual prediction by the LightGBM model, indicating how these changes might influence the campaign's likelihood of securing funding.

\begin{figure}[H]
\centering
\includegraphics[width=\textwidth]{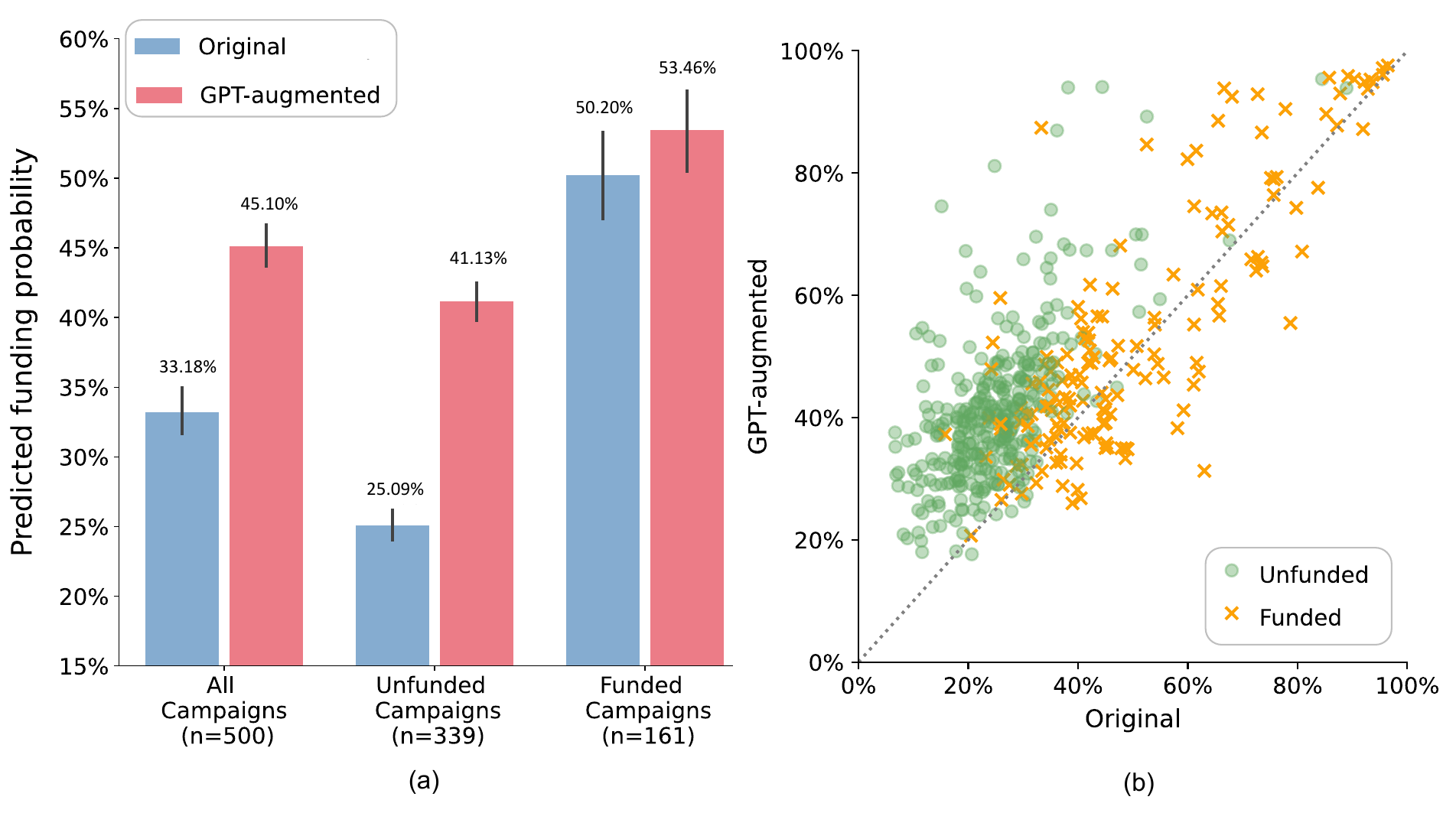}
\caption{Predicted Funding Probability of Original and GPT-augmented Campaign Descriptions in Offline Simulation. 
Subplot (a): the average predicted funding probability of all 500 campaigns significantly increased from 33.18\% to 45.10\%, with the funding probability of the originally unfunded campaigns rising from 25.09\% to 41.13\% (with error bars representing standard errors). Subplot (b): a scatterplot comparing the predicted likelihood of funding before and after ChatGPT augmentation of all 500 campaigns. Originally unfunded campaigns received a higher boosting in the likelihood of funding than originally funded campaigns.}
\label{fig:simulation}
\end{figure}

Our results show that 92\% campaigns are more likely to be funded with the GPT-revised description than with its original description. On average, as shown in Figure \ref{fig:simulation},  the likelihood of fundraising success lifts from 33.2\% before AI augmentation to 45.1\% afterwards, marking a notable 11.9\% average increase (and 12\% , $p<.01$ after controlling for the length of the description -- a robustness check due to the extended text length with GPT augmentation as shown in Table \ref{tab:robust_check_three}).

Which campaigns benefit more from the ChatGPT augmentation? In other words, does revising the campaign description through ChatGPT diminish or amplify the disparity between campaigns that were originally funded and those that were not? 
As shown in Figure \ref{fig:simulation}, campaigns that had failed to secure any fund with their original descriptions could benefit more from the revisions by ChatGPT-4 (the likelihood of fundraising increases from 25.1\% to 41.1\%) than those who had received fund with their original descriptions (the likelihood of fundraising has a smaller increase from 50.2\% to 53.5\%). This effect is particularly desirable, as campaigns organized by female or in low socio-economic regions are less likely to be funded (Figures~\ref{fig:overview} and \ref{fig:equity}), and our result underscores the potential of using AI tools like ChatGPT to enhance equity in crowdfunding success by strategically augmenting campaign descriptions. Further analysis confirms that campaigns organized by female and campaigns from cities with lower education levels on average experience a significantly greater boost in the likelihood of funding when they incorporate expressions of gratitude towards donors, as augmented by ChatGPT (see Table \ref{tab:social_inequity} in SI). 

Additional regression analyses controlling for text length confirm that all these effects remain significant (see Table \ref{tab:robust_check_three}), which rules out alternative hypotheses that the observed differences arise merely from adding more words into the description.

\subsection{Online Experiment}
We complement the offline simulation analysis and validate the effectiveness of AI augmentation, with a randomized online experiment to test whether the GPT-revised campaign descriptions are more preferable to humans (pre-registered on AEA RCT Registry \cite{ye2023ai}).

We randomly selected 16 campaigns from the simulation analysis via stratified sampling (\ref{sec:exp_sample_selection}).  
To control for the length effect, 
we prompted ChatGPT-4 to paraphrase and extend the original description (without introducing additional information), referred to as \emph{GPT-extended}, to match the length to the revised version that implements the three insights (referred to as \emph{GPT-augmented}). For each campaign, we generated pairwise comparisons among these three conditions (original, GPT-augmented, and GPT-extended).  

\begin{figure}[!h]
\centering
\includegraphics[width=0.6\textwidth]{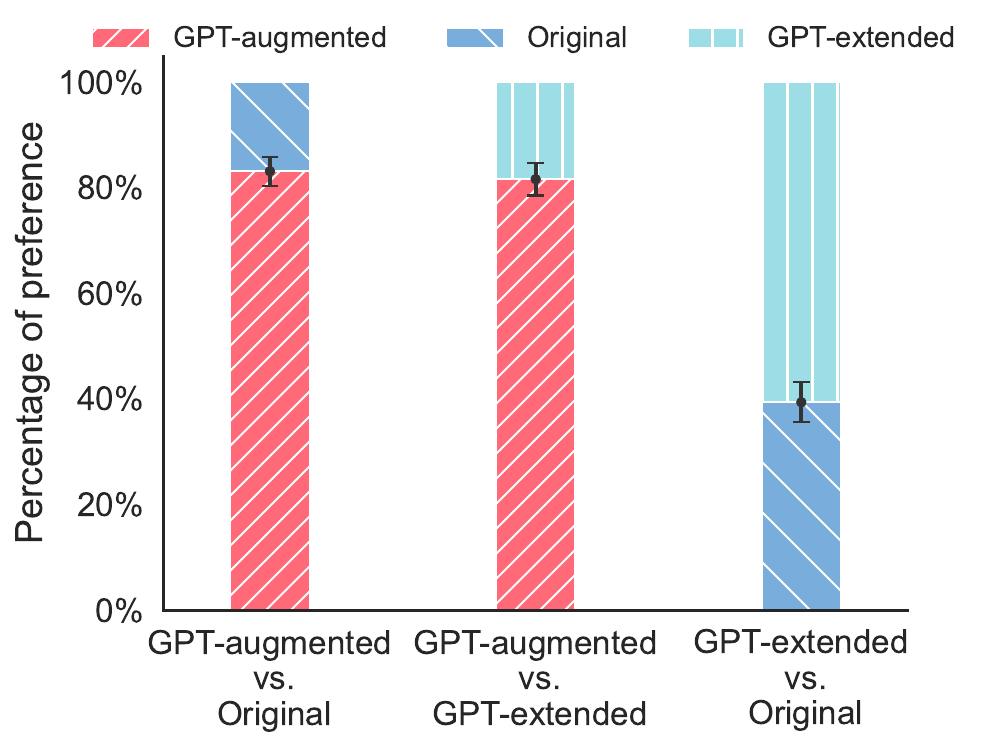}
\caption{Proportions of participant preferences between pairwise comparisons of GPT-augmented, GPT-extended, and original campaign descriptions, with error bars representing standard errors. 83\% (or 82\%) participants preferred GPT-augmented campaigns over the original (or GPT-extended) version, and GPT-extended campaigns are favored by 61\% participants when compared to the originals.}
\label{fig:experiment_result_summary}
\end{figure}

We recruited 263 participants from the Prolific survey platform \cite{peer2022data}. Each participant was randomly assigned to two paired comparisons, and they are prompted to indicate their own preference as well as their belief of most people's preference in making a donation to either of the paired versions of campaign descriptions, presented in random order. 
A detailed discussion of our experimental design, summary statistics of our recruited participants, and a copy of key instruments of our questionnaire are included in {\ref{si:experiment}.

As shown in Figure \ref{fig:experiment_result_summary}, 83\% of the participants preferred GPT-augmented campaign descriptions over their original versions. 82\% favored the GPT-augmented versions over the GPT-extended ones. In addition, the GPT-extended versions were preferred by 61\% of the participants over the original descriptions, indicating the text length effect. Logistic regression leveraging the discrete choice model (see Table \ref{tab:survey_logistic}) further reveals that the GPT-augmented descriptions are 36\% ($p<.01$) more likely to receive a donation compared to the original version, while the GPT-extended versions are 7\% ($p<.01$) more likely to receive a donation than the originals. GPT-augmented descriptions significantly outperform the GPT-extended versions ($p<.01$, Wald test), underscoring that the effectiveness of implementing the three strategies, rather than merely increasing the length of the description.



In a follow-up survey to explain their choices, 30.5\% of the participants explicitly cited the availability of a matching grant as a key factor in their decision making. 33.9\% of the participants acknowledged their choice were affected by the fundraisers' expression of gratitude towards the donors. Another 5.5\% highlighted the urgency of the need as one reason behind their choice. For example, the participants mentioned that 
``\textit{the fact that they add the matching grant makes me want to come alongside and see them attain their goal}''; ``\textit{Campaigns that are personable and include expressions of gratitude can create a stronger connection with potential backers}''; ``\textit{I think most people hear desperation and they want to help}''.  The qualitative evidence confirms that the insights discovered by our machine learning models are successfully delivered in the revision made by ChatGPT. While the length of the description is positively associated with funding success, the feedback from participants is mixed: while 28.9\% of participants gave credits to the effort of writing longer descriptions (``\textit{A longer description will make people appreciate the effort, thus they will be more willing to donate}''), 47.9\% mentioned that they preferred shorter and more concise descriptions, as ``\textit{People don't want to read a book. Keep it short and concise}''.


\section{Discussion} 
\label{sec:discussion}
Unlocking the crowdfunding success for small businesses is of unprecedented importance, particularly in light of the escalating federal debt and the financial constraints faced by many states, which hinder them to meet all their fiscal obligations \cite{truth_2023}. However, optimizing the design and operation of crowdfunding campaigns imposes significant knowledge, experience, and resource barriers on small businesses. 
Utilizing the advancements in large language models and generative AI, our research demonstrates that the success rate of crowdfunding campaigns can be substantially enhanced. This is achieved by applying a few generally implementable, yet effective strategies to refine the narratives in the campaign descriptions before the campaign launch. Our results suggest that even minor adjustments in campaign narratives, facilitated by AI, can have a profound and equitable impact on fundraising outcomes, particularly benefiting small businesses from low socioeconomic regions that might otherwise struggle to attract financial support.

Specifically, our findings indicate that simply including a mention of a matching grant in the campaign description can increase the probability of receiving donations by 5\%. This insight is particularly crucial in light of the mixed results from previous studies, which have shown that while matching grants sometimes boost donation likelihood~\cite{chen2005online,eckel2006subsidizing,karlan2007does}, there are instances where they can have a negative impact, especially when the example matching amount is high (like using \$25:\$75 to illustrate the 1:3 donation-matching rate) or when the match rate is less than 1:1 ~\cite{karlan2011small}.
Until our study, the specific impact of a fixed \$500 match in an online context on charitable giving was unclear. Our research has revealed the substantial impact of a fixed \$500 match, suggesting that crowdfunding platforms and governments could leverage this fixed-amount matching mechanism to enhance fundraising efforts for small businesses.

Our findings reveal a significantly positive influence of expressing gratitude on fundraising outcomes. While a thank-you note might seem inconsequential, it proves to be surprisingly effective. Interestingly, only about 62.86\% of the fundraising campaigns in our dataset included expressions of gratitude in their descriptions. This finding enriches the existing literature on the role of gratitude in prosocial behavior and relationship management (e.g., \citenum{grant2010little, ma2017does}) by providing empirical evidence that a simple expression of thanks can be highly effective in the context of online crowdfunding. 

Our research reveals that articulating the urgency of funding significantly enhances the likelihood of fundraising success. This aligns with previous literature recognizing the critical role of urgency in human decision-making and the preliminary findings via lexicon-based approach in crowdfunding \cite{yazdani2024mis}. As a campaign's deadline draws near, the sense of urgency created by this time-to-closure pressure positively influences charitable giving decisions in online crowdfunding scenarios \cite{kamatham2021effect}. This phenomenon can be attributed to how our brain reacts to deadline-induced pressure: it reduces the threshold of accumulated perceptual evidence required to make a decision \cite{murphy2016global}. Our study adds to this body of literature by examining the impact of urgency triggered by a different type of stimuli. We discovered that conveying urgency within campaign descriptions positively affects donation decisions on GoFundMe, a platform where an official fundraising closure deadline is typically not present. This highlights the importance of how urgency is communicated, even in the absence of a fixed deadline.

These three AI-identified strategies can be applied to a wide range of campaigns. Both the simulation analysis and the online experiment reveal that implementing these strategies via an off-the-shelf large language model, with just a few prompts, can notably increase fundraising success. Additionally, for small businesses that may not have experience with ChatGPT or similar tools, these adjustments can also be carried out manually without needing specific domain knowledge or technical skills. These strategies are particularly advantageous for campaigns that were initially unsuccessful, offering a significant opportunity to improve their outcomes. 

Our research also uncovers insights that could be further validated in field experiments, provided there are suitable subjects and appropriate interventions available. For instance, we find that a small business is more likely to secure funding if it justifies the requested budget in its description. This includes detailing why a specific funding amount is sought, such as to cover expenses like rent or employee salaries, and explaining why these funds are necessary at this particular time. Other suggestions for a small business fundraising campaign include detailing its status quo in light of its own past circumstances but avoid drawing explicit comparisons to peer businesses, writing longer descriptions with simple and concrete language, and avoiding ``spammy'' expressions that may compromise the credibility of the campaign. 

Our findings show that donors are less likely to fund new businesses. This may be connected to risk-aversion theory \cite{menezes1970theory, zhang2014origin} and prior literature on fundraising for startups \cite{angerer2018risk}. This suggests that compared to aiding the sustainability of existing businesses, supporting new businesses and startups poses an even greater challenge. In light of this, there is a need for the creation of specialized policies or mechanisms designed specifically for new businesses and startups, particularly during external challenges such as a pandemic.

In all, the progression of generative AI offers remarkable capabilities for understanding, analyzing, and advising on textual communications. These capabilities present an unprecedented opportunity to optimize descriptions of crowdfunding campaigns, which are typically easy to modify but challenging to strategize effectively. Our study marks a pioneering effort in aiding small businesses to strategically plan their fundraising campaigns with the aid of an AI co-pilot, guiding them through the entire process – from conceptualizing strategies and making counterfactual forecasts to actually implementing these strategies.  

While our data is sourced from a prominent crowdfunding platform during a specific external shock, it's crucial to extend this analysis to other platforms and different time periods. Our findings should not be considered universally applicable to all crowdfunding scenarios. Rather, they should be seen as an initial demonstration of how to employ generative AI and explainable machine learning to enhance the effectiveness and equity of crowdfunding campaigns.






\newpage

\bibliography{sn-bibliography}  

\begin{thebibliography}{10}
\expandafter\ifx\csname url\endcsname\relax
  \def\url#1{\burl{#1}}\fi
\expandafter\ifx\csname urlprefix\endcsname\relax\def\urlprefix{URL }\fi
\providecommand{\bibinfo}[2]{#2}
\providecommand{\eprint}[2][]{\url{#2}}
\providecommand{\doi}[1]{\url{https://doi.org/#1}}
\bibcommenthead

\bibitem{forbes_2022}
\bibinfo{author}{Rowinski, M.}
\newblock \bibinfo{title}{How small businesses drive the american economy}.
\newblock \emph{\bibinfo{journal}{Forbes}}  (\bibinfo{year}{2022}).
\newblock \urlprefix\url{https://www.forbes.com/sites/forbesbusinesscouncil/2022/03/25/how-small-businesses-drive-the-american-economy/?sh=442ad0df4169}.

\bibitem{sba_2023}
\bibinfo{author}{Office, O.~A.}
\newblock \bibinfo{title}{Frequently asked questions about small business 2023}.
\newblock \emph{\bibinfo{journal}{The U.S. Small Business Administration}}  (\bibinfo{year}{2023}).
\newblock \urlprefix\url{https://advocacy.sba.gov/2023/03/07/frequently-asked-questions-about-small-business-2023/#:~:text=Most%20businesses%20are%20small%2D%2099.9,46.4%25%20of%20private%20sector%20employees.}

\bibitem{welsh1981small}
\bibinfo{author}{Welsh, J.~A.}
\newblock \bibinfo{title}{A small business is not a little big business}.
\newblock \emph{\bibinfo{journal}{Harvard business review}} \bibinfo{pages}{July--August} (\bibinfo{year}{1981}).

\bibitem{sahin2011small}
\bibinfo{author}{Sahin, A.}, \bibinfo{author}{Kitao, S.}, \bibinfo{author}{Cororaton, A.} \& \bibinfo{author}{Laiu, S.}
\newblock \bibinfo{title}{Why small businesses were hit harder by the recent recession}.
\newblock \emph{\bibinfo{journal}{Current Issues in Economics and Finance}} \textbf{\bibinfo{volume}{17}} (\bibinfo{year}{2011}).

\bibitem{duygan2015financing}
\bibinfo{author}{Duygan-Bump, B.}, \bibinfo{author}{Levkov, A.} \& \bibinfo{author}{Montoriol-Garriga, J.}
\newblock \bibinfo{title}{Financing constraints and unemployment: Evidence from the great recession}.
\newblock \emph{\bibinfo{journal}{Journal of Monetary Economics}} \textbf{\bibinfo{volume}{75}}, \bibinfo{pages}{89--105} (\bibinfo{year}{2015}).

\bibitem{tierney1997business}
\bibinfo{author}{Tierney, K.~J.}
\newblock \bibinfo{title}{Business impacts of the northridge earthquake}.
\newblock \emph{\bibinfo{journal}{Journal of Contingencies and crisis management}} \textbf{\bibinfo{volume}{5}}, \bibinfo{pages}{87--97} (\bibinfo{year}{1997}).

\bibitem{bartik2020impact}
\bibinfo{author}{Bartik, A.~W.} \emph{et~al.}
\newblock \bibinfo{title}{The impact of covid-19 on small business outcomes and expectations}.
\newblock \emph{\bibinfo{journal}{Proceedings of the national academy of sciences}} \textbf{\bibinfo{volume}{117}}, \bibinfo{pages}{17656--17666} (\bibinfo{year}{2020}).

\bibitem{fairlie2023were}
\bibinfo{author}{Fairlie, R.}, \bibinfo{author}{Fossen, F.~M.}, \bibinfo{author}{Johnsen, R.} \& \bibinfo{author}{Droboniku, G.}
\newblock \bibinfo{title}{Were small businesses more likely to permanently close in the pandemic?}
\newblock \emph{\bibinfo{journal}{Small Business Economics}} \textbf{\bibinfo{volume}{60}}, \bibinfo{pages}{1613--1629} (\bibinfo{year}{2023}).

\bibitem{forbes_2023}
\bibinfo{author}{Center, B.}
\newblock \bibinfo{title}{With vc money slowing, crowdfunding is increasing—here's how to have a successful campaign}.
\newblock \emph{\bibinfo{journal}{Forbes}}  (\bibinfo{year}{2023}).
\newblock \urlprefix\url{https://www.forbes.com/sites/forbesbusinesscouncil/2023/09/08/with-vc-money-slowing-crowdfunding-is-increasing-heres-how-to-have-a-successful-campaign/?sh=7569c6045f1f}.

\bibitem{igra2021crowdfunding}
\bibinfo{author}{Igra, M.}, \bibinfo{author}{Kenworthy, N.}, \bibinfo{author}{Luchsinger, C.} \& \bibinfo{author}{Jung, J.-K.}
\newblock \bibinfo{title}{Crowdfunding as a response to covid-19: Increasing inequities at a time of crisis}.
\newblock \emph{\bibinfo{journal}{Social Science \& Medicine}} \textbf{\bibinfo{volume}{282}}, \bibinfo{pages}{114105} (\bibinfo{year}{2021}).

\bibitem{chen2005online}
\bibinfo{author}{Chen, Y.}, \bibinfo{author}{Li, X.} \& \bibinfo{author}{MacKie-Mason, J.~K.}
\newblock \bibinfo{title}{Online fund-raising mechanisms: A field experiment}.
\newblock \emph{\bibinfo{journal}{Contributions in Economic Analysis \& Policy}} \textbf{\bibinfo{volume}{5}} (\bibinfo{year}{2005}).

\bibitem{karlan2007does}
\bibinfo{author}{Karlan, D.} \& \bibinfo{author}{List, J.~A.}
\newblock \bibinfo{title}{Does price matter in charitable giving? evidence from a large-scale natural field experiment}.
\newblock \emph{\bibinfo{journal}{American Economic Review}} \textbf{\bibinfo{volume}{97}}, \bibinfo{pages}{1774--1793} (\bibinfo{year}{2007}).

\bibitem{falk2007gift}
\bibinfo{author}{Falk, A.}
\newblock \bibinfo{title}{Gift exchange in the field}.
\newblock \emph{\bibinfo{journal}{Econometrica}} \textbf{\bibinfo{volume}{75}}, \bibinfo{pages}{1501--1511} (\bibinfo{year}{2007}).

\bibitem{alpizar2008anonymity}
\bibinfo{author}{Alpizar, F.}, \bibinfo{author}{Carlsson, F.} \& \bibinfo{author}{Johansson-Stenman, O.}
\newblock \bibinfo{title}{Anonymity, reciprocity, and conformity: Evidence from voluntary contributions to a national park in costa rica}.
\newblock \emph{\bibinfo{journal}{Journal of Public Economics}} \textbf{\bibinfo{volume}{92}}, \bibinfo{pages}{1047--1060} (\bibinfo{year}{2008}).

\bibitem{chao2017demotivating}
\bibinfo{author}{Chao, M.}
\newblock \bibinfo{title}{Demotivating incentives and motivation crowding out in charitable giving}.
\newblock \emph{\bibinfo{journal}{Proceedings of the National Academy of Sciences}} \textbf{\bibinfo{volume}{114}}, \bibinfo{pages}{7301--7306} (\bibinfo{year}{2017}).

\bibitem{rhue2018emotional}
\bibinfo{author}{Rhue, L.} \& \bibinfo{author}{Robert, L.~P.}
\newblock \bibinfo{title}{Emotional delivery in pro-social crowdfunding success}.
\newblock \emph{\bibinfo{journal}{Extended Abstracts of the 2018 CHI Conference on Human Factors in Computing Systems}} \bibinfo{pages}{1--6} (\bibinfo{year}{2018}).

\bibitem{zhang2021contributes}
\bibinfo{author}{Zhang, X.}, \bibinfo{author}{Lyu, H.} \& \bibinfo{author}{Luo, J.}
\newblock \bibinfo{title}{What contributes to a crowdfunding campaign's success? evidence and analyses from gofundme data}.
\newblock \emph{\bibinfo{journal}{Journal of Social Computing}} \textbf{\bibinfo{volume}{2}}, \bibinfo{pages}{183--192} (\bibinfo{year}{2021}).

\bibitem{du2021predicting}
\bibinfo{author}{Du, Q.}, \bibinfo{author}{Li, J.}, \bibinfo{author}{Du, Y.}, \bibinfo{author}{Wang, G.~A.} \& \bibinfo{author}{Fan, W.}
\newblock \bibinfo{title}{Predicting crowdfunding project success based on backers' language preferences}.
\newblock \emph{\bibinfo{journal}{Journal of the Association for Information Science and Technology}} \textbf{\bibinfo{volume}{72}}, \bibinfo{pages}{1558--1574} (\bibinfo{year}{2021}).

\bibitem{list2021experimental}
\bibinfo{author}{List, J.~A.}, \bibinfo{author}{Murphy, J.~J.}, \bibinfo{author}{Price, M.~K.} \& \bibinfo{author}{James, A.~G.}
\newblock \bibinfo{title}{An experimental test of fundraising appeals targeting donor and recipient benefits}.
\newblock \emph{\bibinfo{journal}{Nature Human Behaviour}} \textbf{\bibinfo{volume}{5}}, \bibinfo{pages}{1339--1348} (\bibinfo{year}{2021}).

\bibitem{markowitz2021predictive}
\bibinfo{author}{Markowitz, D.~M.} \& \bibinfo{author}{Shulman, H.~C.}
\newblock \bibinfo{title}{The predictive utility of word familiarity for online engagements and funding}.
\newblock \emph{\bibinfo{journal}{Proceedings of the National Academy of Sciences}} \textbf{\bibinfo{volume}{118}}, \bibinfo{pages}{e2026045118} (\bibinfo{year}{2021}).

\bibitem{gilardi2023chatgpt}
\bibinfo{author}{Gilardi, F.}, \bibinfo{author}{Alizadeh, M.} \& \bibinfo{author}{Kubli, M.}
\newblock \bibinfo{title}{Chatgpt outperforms crowd-workers for text-annotation tasks}.
\newblock \emph{\bibinfo{journal}{Proceedings of the National Academy of Sciences}} \textbf{\bibinfo{volume}{120}} (\bibinfo{year}{2023}).

\bibitem{noy2023experimental}
\bibinfo{author}{Noy, S.} \& \bibinfo{author}{Zhang, W.}
\newblock \bibinfo{title}{Experimental evidence on the productivity effects of generative artificial intelligence}.
\newblock \emph{\bibinfo{journal}{Science}} \textbf{\bibinfo{volume}{381}}, \bibinfo{pages}{187--192} (\bibinfo{year}{2023}).

\bibitem{usnews_2023}
\bibinfo{author}{Duggan, W.}
\newblock \bibinfo{title}{7 best crowdfunding platforms}.
\newblock \emph{\bibinfo{journal}{The U.S. News}}  (\bibinfo{year}{2023}).
\newblock \urlprefix\url{https://money.usnews.com/investing/articles/best-crowdfunding-platforms}.

\bibitem{zhou2018project}
\bibinfo{author}{Zhou, M.}, \bibinfo{author}{Lu, B.}, \bibinfo{author}{Fan, W.} \& \bibinfo{author}{Wang, G.~A.}
\newblock \bibinfo{title}{Project description and crowdfunding success: an exploratory study}.
\newblock \emph{\bibinfo{journal}{Information Systems Frontiers}} \textbf{\bibinfo{volume}{20}}, \bibinfo{pages}{259--274} (\bibinfo{year}{2018}).

\bibitem{frey2004social}
\bibinfo{author}{Frey, B.~S.} \& \bibinfo{author}{Meier, S.}
\newblock \bibinfo{title}{Social comparisons and pro-social behavior: Testing “conditional cooperation” in a field experiment}.
\newblock \emph{\bibinfo{journal}{American economic review}} \textbf{\bibinfo{volume}{94}}, \bibinfo{pages}{1717--1722} (\bibinfo{year}{2004}).

\bibitem{tan2016winning}
\bibinfo{author}{Tan, C.}, \bibinfo{author}{Niculae, V.}, \bibinfo{author}{Danescu-Niculescu-Mizil, C.} \& \bibinfo{author}{Lee, L.}
\newblock \bibinfo{title}{Winning arguments: Interaction dynamics and persuasion strategies in good-faith online discussions}.
\newblock \emph{\bibinfo{journal}{Proceedings of the 25th international conference on world wide web}} \bibinfo{pages}{613--624} (\bibinfo{year}{2016}).

\bibitem{brysbaert2014concreteness}
\bibinfo{author}{Brysbaert, M.}, \bibinfo{author}{Warriner, A.~B.} \& \bibinfo{author}{Kuperman, V.}
\newblock \bibinfo{title}{Concreteness ratings for 40 thousand generally known english word lemmas}.
\newblock \emph{\bibinfo{journal}{Behavior research methods}} \textbf{\bibinfo{volume}{46}}, \bibinfo{pages}{904--911} (\bibinfo{year}{2014}).

\bibitem{warriner2013norms}
\bibinfo{author}{Warriner, A.~B.}, \bibinfo{author}{Kuperman, V.} \& \bibinfo{author}{Brysbaert, M.}
\newblock \bibinfo{title}{Norms of valence, arousal, and dominance for 13,915 english lemmas}.
\newblock \emph{\bibinfo{journal}{Behavior research methods}} \textbf{\bibinfo{volume}{45}}, \bibinfo{pages}{1191--1207} (\bibinfo{year}{2013}).

\bibitem{wang2015making}
\bibinfo{author}{Wang, B.}, \bibinfo{author}{Zubiaga, A.}, \bibinfo{author}{Liakata, M.} \& \bibinfo{author}{Procter, R.}
\newblock \bibinfo{title}{Making the most of tweet-inherent features for social spam detection on twitter}.
\newblock \emph{\bibinfo{journal}{arXiv preprint arXiv:1503.07405}}  (\bibinfo{year}{2015}).

\bibitem{ke2017lightgbm}
\bibinfo{author}{Ke, G.} \emph{et~al.}
\newblock \bibinfo{title}{Lightgbm: A highly efficient gradient boosting decision tree}.
\newblock \emph{\bibinfo{journal}{Advances in neural information processing systems}} \textbf{\bibinfo{volume}{30}} (\bibinfo{year}{2017}).

\bibitem{OpenAI}
\bibinfo{title}{{OpenAI API}} (\bibinfo{year}{2023}).
\newblock \urlprefix\url{https://openai.com/blog/openai-api}.

\bibitem{duan2020entrepreneurs}
\bibinfo{author}{Duan, Y.}, \bibinfo{author}{Hsieh, T.-S.}, \bibinfo{author}{Wang, R.~R.} \& \bibinfo{author}{Wang, Z.}
\newblock \bibinfo{title}{Entrepreneurs' facial trustworthiness, gender, and crowdfunding success}.
\newblock \emph{\bibinfo{journal}{Journal of Corporate Finance}} \textbf{\bibinfo{volume}{64}}, \bibinfo{pages}{101693} (\bibinfo{year}{2020}).

\bibitem{zhu2022proximal}
\bibinfo{author}{Zhu, X.}
\newblock \bibinfo{title}{Proximal language predicts crowdfunding success: Behavioral and experimental evidence}.
\newblock \emph{\bibinfo{journal}{Computers in Human Behavior}} \textbf{\bibinfo{volume}{131}}, \bibinfo{pages}{107213} (\bibinfo{year}{2022}).

\bibitem{shang2009field}
\bibinfo{author}{Shang, J.} \& \bibinfo{author}{Croson, R.}
\newblock \bibinfo{title}{A field experiment in charitable contribution: The impact of social information on the voluntary provision of public goods}.
\newblock \emph{\bibinfo{journal}{The economic journal}} \textbf{\bibinfo{volume}{119}}, \bibinfo{pages}{1422--1439} (\bibinfo{year}{2009}).

\bibitem{ye2023ai}
\bibinfo{author}{Ye, T.} \emph{et~al.}
\newblock \bibinfo{title}{Ai-augmented crowdfunding campaign: An online experiment.}
\newblock \emph{\bibinfo{journal}{AEA RCT Registry}} .

\bibitem{peer2022data}
\bibinfo{author}{Peer, E.}, \bibinfo{author}{Rothschild, D.}, \bibinfo{author}{Gordon, A.}, \bibinfo{author}{Evernden, Z.} \& \bibinfo{author}{Damer, E.}
\newblock \bibinfo{title}{Data quality of platforms and panels for online behavioral research}.
\newblock \emph{\bibinfo{journal}{Behavior Research Methods}} \bibinfo{pages}{1} (\bibinfo{year}{2022}).

\bibitem{truth_2023}
\bibinfo{author}{Willard, J.}
\newblock \bibinfo{title}{Financial state of the states 2023}.
\newblock \emph{\bibinfo{journal}{Truth in Accounting}}  (\bibinfo{year}{2023}).
\newblock \urlprefix\url{https://www.truthinaccounting.org/news/detail/financial-state-of-the-states-2023}.

\bibitem{eckel2006subsidizing}
\bibinfo{author}{Eckel, C.~C.} \& \bibinfo{author}{Grossman, P.~J.}
\newblock \bibinfo{title}{Subsidizing charitable giving with rebates or matching: Further laboratory evidence}.
\newblock \emph{\bibinfo{journal}{Southern Economic Journal}} \textbf{\bibinfo{volume}{72}}, \bibinfo{pages}{794--807} (\bibinfo{year}{2006}).

\bibitem{karlan2011small}
\bibinfo{author}{Karlan, D.}, \bibinfo{author}{List, J.~A.} \& \bibinfo{author}{Shafir, E.}
\newblock \bibinfo{title}{Small matches and charitable giving: Evidence from a natural field experiment}.
\newblock \emph{\bibinfo{journal}{Journal of Public Economics}} \textbf{\bibinfo{volume}{95}}, \bibinfo{pages}{344--350} (\bibinfo{year}{2011}).

\bibitem{grant2010little}
\bibinfo{author}{Grant, A.~M.} \& \bibinfo{author}{Gino, F.}
\newblock \bibinfo{title}{A little thanks goes a long way: Explaining why gratitude expressions motivate prosocial behavior.}
\newblock \emph{\bibinfo{journal}{Journal of personality and social psychology}} \textbf{\bibinfo{volume}{98}}, \bibinfo{pages}{946} (\bibinfo{year}{2010}).

\bibitem{ma2017does}
\bibinfo{author}{Ma, L.~K.}, \bibinfo{author}{Tunney, R.~J.} \& \bibinfo{author}{Ferguson, E.}
\newblock \bibinfo{title}{Does gratitude enhance prosociality?: A meta-analytic review.}
\newblock \emph{\bibinfo{journal}{Psychological bulletin}} \textbf{\bibinfo{volume}{143}}, \bibinfo{pages}{601} (\bibinfo{year}{2017}).

\bibitem{yazdani2024mis}
\bibinfo{author}{Yazdani, E.}, \bibinfo{author}{Chakravarty, A.} \& \bibinfo{author}{Inman, J.}
\newblock \bibinfo{title}{(mis) alignment between facial and textual emotions and its effects on donors engagement behavior in online crowdsourcing platforms}.
\newblock \emph{\bibinfo{journal}{Journal of the Academy of Marketing Science}} \bibinfo{pages}{1--21} (\bibinfo{year}{2024}).

\bibitem{kamatham2021effect}
\bibinfo{author}{Kamatham, S.~H.}, \bibinfo{author}{Pahwa, P.}, \bibinfo{author}{Jiang, J.} \& \bibinfo{author}{Kumar, N.}
\newblock \bibinfo{title}{Effect of appeal content on fundraising success and donor behavior}.
\newblock \emph{\bibinfo{journal}{Journal of Business Research}} \textbf{\bibinfo{volume}{125}}, \bibinfo{pages}{827--839} (\bibinfo{year}{2021}).

\bibitem{murphy2016global}
\bibinfo{author}{Murphy, P.~R.}, \bibinfo{author}{Boonstra, E.} \& \bibinfo{author}{Nieuwenhuis, S.}
\newblock \bibinfo{title}{Global gain modulation generates time-dependent urgency during perceptual choice in humans}.
\newblock \emph{\bibinfo{journal}{Nature communications}} \textbf{\bibinfo{volume}{7}}, \bibinfo{pages}{13526} (\bibinfo{year}{2016}).

\bibitem{menezes1970theory}
\bibinfo{author}{Menezes, C.~F.} \& \bibinfo{author}{Hanson, D.~L.}
\newblock \bibinfo{title}{On the theory of risk aversion}.
\newblock \emph{\bibinfo{journal}{International Economic Review}} \bibinfo{pages}{481--487} (\bibinfo{year}{1970}).

\bibitem{zhang2014origin}
\bibinfo{author}{Zhang, R.}, \bibinfo{author}{Brennan, T.~J.} \& \bibinfo{author}{Lo, A.~W.}
\newblock \bibinfo{title}{The origin of risk aversion}.
\newblock \emph{\bibinfo{journal}{Proceedings of the National Academy of Sciences}} \textbf{\bibinfo{volume}{111}}, \bibinfo{pages}{17777--17782} (\bibinfo{year}{2014}).

\bibitem{angerer2018risk}
\bibinfo{author}{Angerer, M.}, \bibinfo{author}{Niemand, T.}, \bibinfo{author}{Kraus, S.} \& \bibinfo{author}{Thies, F.}
\newblock \bibinfo{title}{Risk-reducing options in crowdinvesting: An experimental study}.
\newblock \emph{\bibinfo{journal}{Journal of Small Business Strategy (archive only)}} \textbf{\bibinfo{volume}{28}}, \bibinfo{pages}{1--17} (\bibinfo{year}{2018}).

\bibitem{mohammad2013crowdsourcing}
\bibinfo{author}{Mohammad, S.~M.} \& \bibinfo{author}{Turney, P.~D.}
\newblock \bibinfo{title}{Crowdsourcing a word--emotion association lexicon}.
\newblock \emph{\bibinfo{journal}{Computational intelligence}} \textbf{\bibinfo{volume}{29}}, \bibinfo{pages}{436--465} (\bibinfo{year}{2013}).

\bibitem{chen2023investigating}
\bibinfo{author}{Chen, Y.}, \bibinfo{author}{Zhou, S.}, \bibinfo{author}{Jin, W.} \& \bibinfo{author}{Chen, S.}
\newblock \bibinfo{title}{Investigating the determinants of medical crowdfunding performance: A signaling theory perspective}.
\newblock \emph{\bibinfo{journal}{Internet Research}} \textbf{\bibinfo{volume}{33}}, \bibinfo{pages}{1134--1156} (\bibinfo{year}{2023}).

\bibitem{ullah2020gender}
\bibinfo{author}{Ullah, S.} \& \bibinfo{author}{Zhou, Y.}
\newblock \bibinfo{title}{Gender, anonymity and team: What determines crowdfunding success on kickstarter}.
\newblock \emph{\bibinfo{journal}{Journal of Risk and Financial Management}} \textbf{\bibinfo{volume}{13}}, \bibinfo{pages}{80} (\bibinfo{year}{2020}).

\bibitem{mueller2021impacts}
\bibinfo{author}{Mueller, J.~T.} \emph{et~al.}
\newblock \bibinfo{title}{Impacts of the covid-19 pandemic on rural america}.
\newblock \emph{\bibinfo{journal}{Proceedings of the National Academy of Sciences}} \textbf{\bibinfo{volume}{118}}, \bibinfo{pages}{2019378118} (\bibinfo{year}{2021}).

\bibitem{united2019american}
\bibinfo{title}{{American Community Survey Data Tables}} (\bibinfo{year}{2019}).
\newblock \urlprefix\url{https://www.census.gov/programs-surveys/acs/data/data-tables.html}.

\bibitem{peng2022speaking}
\bibinfo{author}{Peng, L.}, \bibinfo{author}{Cui, G.}, \bibinfo{author}{Bao, Z.} \& \bibinfo{author}{Liu, S.}
\newblock \bibinfo{title}{Speaking the same language: the power of words in crowdfunding success and failure}.
\newblock \emph{\bibinfo{journal}{Marketing Letters}} \textbf{\bibinfo{volume}{33}}, \bibinfo{pages}{311--323} (\bibinfo{year}{2022}).

\bibitem{ye2020predicting}
\bibinfo{author}{Ye, T.} \emph{et~al.}
\newblock \bibinfo{title}{Predicting individual treatment effects of large-scale team competitions in a ride-sharing economy}.
\newblock \emph{\bibinfo{journal}{Proceedings of the 26th ACM SIGKDD International Conference on Knowledge Discovery \& Data Mining}} \bibinfo{pages}{2368--2377} (\bibinfo{year}{2020}).

\bibitem{wei2022chain}
\bibinfo{author}{Wei, J.} \emph{et~al.}
\newblock \bibinfo{title}{Chain-of-thought prompting elicits reasoning in large language models}.
\newblock \emph{\bibinfo{journal}{Advances in Neural Information Processing Systems}} \textbf{\bibinfo{volume}{35}}, \bibinfo{pages}{24824--24837} (\bibinfo{year}{2022}).

\bibitem{kenworthy2020cross}
\bibinfo{author}{Kenworthy, N.}, \bibinfo{author}{Dong, Z.}, \bibinfo{author}{Montgomery, A.}, \bibinfo{author}{Fuller, E.} \& \bibinfo{author}{Berliner, L.}
\newblock \bibinfo{title}{A cross-sectional study of social inequities in medical crowdfunding campaigns in the united states}.
\newblock \emph{\bibinfo{journal}{PLoS One}} \textbf{\bibinfo{volume}{15}}, \bibinfo{pages}{e0229760} (\bibinfo{year}{2020}).

\bibitem{coleman2009comparison}
\bibinfo{author}{Coleman, S.} \& \bibinfo{author}{Robb, A.}
\newblock \bibinfo{title}{A comparison of new firm financing by gender: evidence from the kauffman firm survey data}.
\newblock \emph{\bibinfo{journal}{Small Business Economics}} \textbf{\bibinfo{volume}{33}}, \bibinfo{pages}{397--411} (\bibinfo{year}{2009}).

\bibitem{bapna2021gender}
\bibinfo{author}{Bapna, S.} \& \bibinfo{author}{Ganco, M.}
\newblock \bibinfo{title}{Gender gaps in equity crowdfunding: Evidence from a randomized field experiment}.
\newblock \emph{\bibinfo{journal}{Management Science}} \textbf{\bibinfo{volume}{67}}, \bibinfo{pages}{2679--2710} (\bibinfo{year}{2021}).

\bibitem{gafni2021gender}
\bibinfo{author}{Gafni, H.}, \bibinfo{author}{Marom, D.}, \bibinfo{author}{Robb, A.} \& \bibinfo{author}{Sade, O.}
\newblock \bibinfo{title}{Gender dynamics in crowdfunding (kickstarter): Evidence on entrepreneurs, backers, and taste-based discrimination}.
\newblock \emph{\bibinfo{journal}{Review of Finance}} \textbf{\bibinfo{volume}{25}}, \bibinfo{pages}{235--274} (\bibinfo{year}{2021}).

\bibitem{greenberg2015leaning}
\bibinfo{author}{Greenberg, J.} \& \bibinfo{author}{Mollick, E.}
\newblock \bibinfo{title}{Leaning in or leaning on? gender, homophily, and activism in crowdfunding}.
\newblock \emph{\bibinfo{journal}{Academy of Management Proceedings}}  (\bibinfo{year}{2015}).

\bibitem{kanze2018we}
\bibinfo{author}{Kanze, D.}, \bibinfo{author}{Huang, L.}, \bibinfo{author}{Conley, M.~A.} \& \bibinfo{author}{Higgins, E.~T.}
\newblock \bibinfo{title}{We ask men to win and women not to lose: Closing the gender gap in startup funding}.
\newblock \emph{\bibinfo{journal}{Academy of Management Journal}} \textbf{\bibinfo{volume}{61}}, \bibinfo{pages}{586--614} (\bibinfo{year}{2018}).

\bibitem{prelec2004bayesian}
\bibinfo{author}{Prelec, D.}
\newblock \bibinfo{title}{A bayesian truth serum for subjective data}.
\newblock \emph{\bibinfo{journal}{science}} \textbf{\bibinfo{volume}{306}}, \bibinfo{pages}{462--466} (\bibinfo{year}{2004}).

\bibitem{oppenheimer2009instructional}
\bibinfo{author}{Oppenheimer, D.~M.}, \bibinfo{author}{Meyvis, T.} \& \bibinfo{author}{Davidenko, N.}
\newblock \bibinfo{title}{Instructional manipulation checks: Detecting satisficing to increase statistical power}.
\newblock \emph{\bibinfo{journal}{Journal of experimental social psychology}} \textbf{\bibinfo{volume}{45}}, \bibinfo{pages}{867--872} (\bibinfo{year}{2009}).

\bibitem{athey2022smiles}
\bibinfo{author}{Athey, S.}, \bibinfo{author}{Karlan, D.}, \bibinfo{author}{Palikot, E.} \& \bibinfo{author}{Yuan, Y.}
\newblock \bibinfo{title}{Smiles in profiles: Improving fairness and efficiency using estimates of user preferences in online marketplaces}.
\newblock \bibinfo{type}{Tech. Rep.}, \bibinfo{institution}{National Bureau of Economic Research} (\bibinfo{year}{2022}).

\end{thebibliography}

\begin{appendices}
\renewcommand{\thesection}{SI}

\newpage
\section{Supplementary Information}
\label{app:SI}

\setcounter{figure}{0}
\renewcommand{\thefigure}{S\arabic{figure}}

\setcounter{table}{0}
\renewcommand{\thetable}{S\arabic{table}}

\subsection {Predicative Analysis}

\subsubsection{Data Collection and Cleaning}
\label{si-sec:data_collection}
In this study, our analysis focuses on investigating crowdfunding for small businesses in the United States. Therefore, we collect all fundraising campaigns that fall into the collection of ``small businesses affected by the coronavirus'' from GoFundMe (GFM), which were created during the most recent pandemic shock from January $22^{nd}$, 2020 to December $31^{st}$, 2020. We filter out non-business campaigns and remove campaigns outside of the U.S., and blank campaigns lacking key information (such as text description, location, and posting date), resulting in a dataset of 12,646 campaigns. In addition, we employ ChatGPT API with the "gpt-4-1106-preview" to filter out campaigns that does not target sponsoring individual small businesses (see prompt in Table \ref{tab:prompt_busi}).\footnote{To ensure the quality of the ChatGPT classification, we random select 100 campaigns and find the independent human evaluation highly consistent with the ChatGPT labels (Cohen's Kappa = $1.0$).} 

\vspace{20pt}

\begin{table}[h]
\centering 
\caption{Prompt for small business validation}
\label{tab:prompt_busi}
\scriptsize
\begin{tabular}{{l}}

\toprule
\\[-2ex] 
\makecell[l]{
\textbf{PROMPT:}
\\
\\[-1ex]
You will be provided with a text delimited by triple quotes. \\
The text comes from a crowdfunding campaign description. \\
It's trying to raise money for a business. \\
You have these following tasks, please output the result in JSON format:\\\\
\textbf{Task 1} : Determine if the campaign is about a small business. \\
Output TRUE or FALSE to field business, and explanation to business\_explanation. \\ If the answer to Task 1 is TRUE, terminate and no need to do the following task. \\If the answer to Task 1 is FALSE, continue to finish the following task. \\
\\ 
\textbf{Task 2} : Then, determine if the campaign is raised by the owner of a business \\ to support its employees. Output TRUE or FALSE to field owner\_support, \\ and explanation to owner\_support\_explanation. }\\
\\[-2ex]
\botrule
\end{tabular}

\end{table}
    
\vspace{20pt}

Our final dataset after data cleaning contains 11,274 campaigns across 50 States in the U.S. The record of each campaign contains the ``story'' (i.e., textual description) and the campaign configuration (such as organizer name, fundraising goal amount, and campaign created date), as well as the time and amount of each donation.

\subsubsection{Feature Engineering}
\label{SI:features}
We construct a total of 168 features to identify the predictive factors related to fundraising success. Table \ref{tab:features} provides an overview of the features in our analysis, categorized into four groups as follows.

\begin{table}[ht]
\centering
\caption{Overview of feature groups}
\small
\label{tab:features}
\begin{tabular}{@{}l|c|l@{}}
\toprule
\textbf{Feature Group} & \textbf{\# of features} & \textbf{Feature Group Descriptions}  \\
\midrule
 Textual description \textsuperscript{\ref{sec:text}.A}& 116 & Textual features (lexicon-based and GPT-generated)).\\
 
 Campaign configuration \textsuperscript{\ref{sec:gfmmeta}.B}  & 4 &  Meta information, e.g., fundraising goal amount. \\

  Level of pandemic shock \textsuperscript{\ref{sec:covid}.C} & 2 & COVID-19 related statistics.\\

  Local demographics \textsuperscript{\ref{sec:acs}.D}& 46 & \makecell[l]{Features from American Community Survey (ACS).}\\ 
 
\bottomrule
\end{tabular}
\end{table}

\noindent\textbf{A. Textual description features}\label{sec:text}

The textual description serves as the primary medium for communicating essential information about the fundraising campaign \cite{zhou2018project}. It addresses critical questions such as why a small business is raising funds, the nature of its services, and the mutual benefits for both the business and its donors \cite{list2021experimental}. 

In this paper, we adopt a lexicon-based approach to represent writing style and sentiment in line with prior research, and we innovatively utilize large language models to capture semantic and pragmatic characteristics from campaign descriptions. Specifically, leveraging prior literature in crowdfunding and small business domain context \cite{list2021experimental, frey2004social}, we generate 11 features using ChatGPT, capturing aspects such as social comparison with peer businesses, plan of funding usage, and urgency highlighted in campaign descriptions. To ensure the quality of ChatGPT outputs, a human evaluator independently labels a sample of 100 randomly selected campaigns. We assess the inter-rater reliability between ChatGPT and human labels using Cohen's kappa for the 11 features, as ordered in the GPT prompt from Task 1 - 11. The kappa values for each feature are as follows: Employees mentioned (0.96), Rent of business mentioned (0.85), Businesses longer than 2 years (0.97), New business (0.77), Match grant mentioned (1.0), Gratitude expressed (0.98), Urgency explained (0.81),  Social Comparison $better \ than \ peers$ (0.80), Self Comparison $worse \ than \ before$ (1.0), Small business specified (0.99), Extrinsic incentive (0.77).

In addition, we develop 105 lexicon-based features, including emotion measurement, text complexity (such as syllables per word and reading grade), as well as text concreteness and dominance. The effects of these features on effective messaging have been confirmed by prior literature in the fields of crowdfunding \cite{markowitz2021predictive, rhue2018emotional, tan2016winning} and other domains \cite{brysbaert2014concreteness, warriner2013norms,wang2015making}. See Table \ref{tab:text_features} for a complete list of textual features grouped by source.

\begin{table}[h]
\centering
\caption{Textual description features}
\label{tab:text_features}
\small
\begin{tabular}{@{}l|c|l@{}}
\toprule

\textbf{Source} & \textbf{\# of features} & \textbf{Method Description}  \\
\midrule
GPT-4 & 11 & \makecell[l]{We adopt the most up-to-date large language model, \\ GPT-4 (gpt-4-1106-preview) to generate 11 
textual features. \\ Prompts used to generate these features are detailed in Table \ref{tab:prompt_feature}. \\ Each task in the prompt refers to one feature.} \\

 \midrule
LIWC-22 & 94 &  We adopt the widely used LIWC-22 dictionary.\footnote{\url{https://www.liwc.app/}}\\
\midrule

\makecell[l]{Other \\lexicon-based \\features}  & 11 & \makecell[l]{
      Concreteness level of words \cite{brysbaert2014concreteness}\\
      Dominance level of words \cite{warriner2013norms}\\
      Flesch-Kincaid grade level\\
      Syllables per word
      \footnote{\url{https://pypi.org/project/textstat/}} \\
      Polarity (from TextBlob package)
      \footnote{\url{https://textblob.readthedocs.io/en/dev/}} \\
      NRC Emotional Lexicon \cite{mohammad2013crowdsourcing} (joy / sadness / positive / negative scores)\\
      Check if the text contains spam words \cite{wang2015making} \\
       Check if the text mentions the names of persons \footnote{\url{https://spacy.io/usage/models}}
}\\

\botrule
\end{tabular}
\end{table}

\setlength\tabcolsep{0pt}
\begin{table}[h]
\centering 
\caption{Prompt for textual feature generation}
\label{tab:prompt_feature}
\scriptsize
\begin{tabular}{{l}}

\toprule
\\[-2ex] 
\makecell[l]{
\textbf{PROMPT:}
\\
You will be provided with a text delimited by triple quotes. The text comes from a crowdfunding \\
campaign description. It's trying to raise money for a business. You have these following tasks, \\
please output the result in JSON format: \\
\\
\textbf{Task 1} : Check whether the author has mentioned that the raised funds is going to help the \\
employees of the business. Output TRUE or FALSE to field [Employee mentioned], \\
and explanation to [employee explanation]. \\
\\[-1ex]
\textbf{Task 2}: Check whether the author has mentioned that the raised funds is going to cover the rent \\
of the business. Output TRUE or FALSE to field [Rent mentioned], and explanation to [rent explanation].\\
\\[-1ex]
\textbf{Task 3}: Determine if the business has a long history.  You are going to check: \\
1. whether the text explicitly suggests that the business has been operating for more than two years.\\
2. whether the text indicating it self as a business with a long history.\\
Note that you are determining if the business has been operated for a long time, but not some \\
professionals working in this business has been working for a long time, because these professionals \\
could have been working in other places.
If you are using individual professionals to make the decision, \\
please make sure it clearly states that this professional has always
been working in the particular business. \\
If either one of the two conditions is true, output TRUE, otherwise output FALSE. \\
Explain why you make such decision. Output the answer to the field  [Business longer than 2 years], \\
and explanation to the  field [long history explanation].\\
\\[-1ex]
\textbf{Task 4}: Determine if the business is newly started (already established but explicitly states it is new).\\
Output TRUE or FALSE to field [New business], and explanation to [new business explanation].\\
\\[-1ex]
\textbf{Task 5}: Determine whether the text mentions that if the author can raise \$500,\\
GoFundMe's Small Business Relief Initiative will match \$500 for the business, or expresses similar \\ meanings,
and explain why you make such decision. Return the result TRUE or FALSE to the \\
field [Match grant mentioned], and the explanation goes to the field [grant explanation].\\
\\[-1ex]
\textbf{Task 6}: Determine whether the text is explicitly expressing gratitude to potential backer, and explain \\
why you think it mentions or not. Return the result TRUE or FALSE goes to the field [Gratitude expressed],\\
and the explanation to the field [gratitude explanation].\\
\\[-1ex]
\textbf{Task 7}:  Determine if the author has mentioned that the need of funds for the business is very urgent \\
in the description. Output TRUE or FALSE to field [Urgency explained], and explanation \\
to [urgency explanation].\\
\\[-1ex]
\textbf{Task 8}: Determine if the provided text includes social comparison and indicates its business
outperforms \\
its peers. The comparisons are usually between the products or the service the business provides, \\
but not limited to this kind. Output TRUE or FALSE to field [Social comparison (better than peers)], \\
and explanation to [social comparison better explanation].\\
\\[-1ex]
\textbf{Task 9}: Determine if the provided text includes self comparison and indicates its business is weaker \\
than before. Output TRUE or FALSE to field [Self comparison (worse than before)], and explanation \\
to [self comparison worse explanation].
\\
\\[-1ex]
\textbf{Task 10} : Check if the campaign has a tag starting with the pound sign \#. If so, output the content \\
of this tag to the field [Tag],
otherwise output ``NO TAG'' to [Tag]. Check if the tag indicates that it is \\
a small business. Output the result as TRUE or 
FALSE to the field [Small Business Specified].\\
\\[-1ex]
\textbf{Task 11}: Determine if the author has mentioned to send some small thank-you gift to potential backers.\\
Notice that the author should be always sending the same gifts no matter how much money is donated. \\
Output TRUE or FALSE to field [Extrinsic incentive], and explanation to [extrinsic incentive explanation].}\\

\botrule
\end{tabular}
\end{table}
\setlength\tabcolsep{6pt}

\noindent\textbf{B. Campaign Configuration Features}\label{sec:gfmmeta}

Campaign configuration may play an important role in fundraising success. To maximize the predictive power of the machine learning models, we include the configuration information that might be associated with the fundraising success, such as the fundraising goal amount \cite{chen2023investigating} and the gender of the fundraiser \cite{ullah2020gender}. Detailed descriptions are shown in Table \ref{tab:gfmmeta}.

\begin{table}[ht]
\centering
\caption{GFM campaign configuration}
\label{tab:gfmmeta}
\small
\begin{tabular}{@{}l|l@{}}
\toprule
\textbf{Feature Name} &  \textbf{Feature Description}  \\
\midrule

GFM beneficiary & If the author of the campaign and the beneficiary is not the same identity. \\
GFM gofundme organize & If the campaign is organized by GoFundMe website.\\
GFM organizer male & If the campaign is organized by a male author.\\
GFM goal amount & The fundraising goal of the campaign in $\$$.\\
\bottomrule
\end{tabular}
\end{table}

\noindent\textbf{C. Level of Pandemic Shock Features}\label{sec:covid}

Since our dataset is collected during the economic recession of the COVID-19 pandemic and its divergent influence on different geographical areas may affect the donation preferences of crowdfunding participants \cite{bartik2020impact, mueller2021impacts}, we control for the severity of the external pandemic shock at the state level by incorporating factors such as the number of new COVID-19 cases within seven days before a campaign's creation date.\footnote{The data are collected from \url{https://covid.cdc.gov/covid-data-tracker/\#datatracker-home}.}



\noindent\textbf{D. Local Demographics Features}\label{sec:acs}

To control for the potential impact of local socio-economic and demographic factors \cite{igra2021crowdfunding}, we select 46 features from 2019 American Community Survey (ACS) \cite{united2019american} to capture local contextual factors such as city income level, business volume, citizen educational level. 
We discuss the details of selected ACS features in Table \ref{tab:acsfeature}.

\newgeometry{left=2.5cm}
\begin{table}[h]
\centering
\vspace{-10mm} 
\caption{Detailed ACS feature groups and feature specifications}
\label{tab:acsfeature}
\small
\begin{tabular}{@{}c|l@{}}
\toprule
\textbf{Feature Group} &  \textbf{Example Features}  \\
\midrule
\textbf{Population} & \makecell[l]{Population per square mile}\\
\midrule
\textbf{Genders} & Percentage of female \\
\midrule
\textbf{Ages} & \makecell[l]{Percentage of persons under 5 years old,\\
Percentage of persons under 18 years old,\\
Percentage of persons over 65 years old \\}\\

\midrule
\textbf{Race} & \makecell[l]{
Percentage of persons of two or more races,\\
White alone not Hispanic or Latino percentage,\\
Asian alone percentage,  Hispanic or Latino percentage,
\\
White alone percentage, Black or African American alone percentage,\\
Native Hawaiian and Other Pacific Islander alone percentage
}\\
\midrule
\textbf{Education} & 
Percentage of population aged 25 years and older with a Bachelor's degree or higher\\
\midrule
\textbf{Household} & \makecell[l]{
Number of housing units, \\
Number of persons per household,
Owner\-occupied housing unit rate,\\
Number of households,
Percentage of persons living in the same house as one year ago,\\
Percentage of households with a computer,\\
Percentage of households with broadband Internet access
\\
}\\
\midrule
\textbf{Income} & \makecell[l]{
Median gross rent, Per capita income in past 12 months,\\
Median household income,
Percentage of persons in poverty,\\
Median value of owner\-occupied housing units,\\
Median selected monthly owner costs with mortgage,\\
Median selected monthly owner costs without mortgage\\
}\\
\midrule
\textbf{Business} & \makecell[l]{Number of all firms,
Number of men\-owned firms, \\
Number of nonminority\-owned firms,\\
Total retail sales,\\
Total retail sales per capita,\\
Total health care and social assistance receipts\/revenue,\\
Business transportation and warehousing receipts\/revenue}\\
\midrule
\textbf{Employment} & \makecell[l]{
Total employer establishments,\\
Total annual payroll,\\
Percentage change in total employment,\\
Total nonemployer establishments,\\
Mean travel time to work for persons aged 16 years and older,
\\Percentage of the civilian labor force aged 16 years and older,\\
Percentage of female in civilian labor force female aged 16 years and older}\\
\midrule
\textbf{Others} & \makecell[l]{Veterans,
Percentage of foreign-born persons,\\
Percentage of persons aged 5 years and older speaking a language other than English at home,\\
Percentage of persons under 65 years with a disability}\\
\botrule
\end{tabular}
\end{table}
\restoregeometry

\subsubsection{Prediction Models and Performances}
\label{si-sec:prediction_model}
We select the Light Gradient-Boosting Machine (LightGBM) model \cite{ke2017lightgbm}, a non-linear tree-based machine learning model of both high computational efficiency and high explainability. This is because our goal of predictive analysis is not only to test the predictive power of the features, but more importantly to discover actionable insights that inform the optimal writing of crowdfunding campaigns. 

Two baselines are considered: a uniform prediction baseline, and a random prediction baseline. The uniform baseline predicts fundraising success based on majority vote (i.e., all campaigns predicted as successfully funded), while the random baseline predicts fundraising success by random sampling from a Bernoulli distribution with a success probability $(p) = \frac{ \text{\# of funded campaigns in the training set}}{\text{total \# }  \text{of campaigns in the training set}}$ . 
\vskip 6pt
We divide the dataset into training, validation, and test sets based on the campaign posting date, following the standard practices in line with time-dependent restrictions, as shown in Table \ref{tab:data_split}.

\begin{table}[h]
\centering 
\caption{Data split}
\label{tab:data_split}
\begin{tabular}{l*{3}{c}}

\toprule
\\[-2ex]
&{Training Set}&{Validation Set}&{Testing Set}\\
\\[-2ex] 
\midrule
\\[-2ex]         
\textbf{\# of Campaigns} & 7391 & 2193 & 1690 \\
\\[-2ex]
\textbf{Starting Date} & 2020-01-22 & 2020-04-01 & 2020-05-01 \\
\\[-2ex]
\textbf{Ending Date} & 2020-03-31 & 2020-04-30 & 2020-12-31 \\

\botrule
\end{tabular}

\end{table}

We rigorously tune the hyper-parameters of the machine learning models based on validation F1-score and report the performance of the models on the test set (fundraising campaigns starting in May, 2020) as shown in Table \ref{tab:lgb}.


According to Table \ref{tab:lgb}, LightGBM significantly outperforms both the uniform baseline (by 36.6\%) and the random baseline (by 48.9\%), indicating the high predictive power of our approach. \\

Additional ablation experiments (Table \ref{tab:lgb_ablation}) suggest that adding the 11 GPT-generated features to the model that only includes all the non-textual features (i.e., campaign configuration, level of pandemic shock, and local demographics), the accuracy of the best-performing LightGBM model improved by 14.6\% (from 59.3\% to 73.9\%). This performance increase demonstrates the effectiveness of ChatGPT-generated features in predicting fundraising success.

\begin{table}[h]
\centering 
\caption{LightGBM model performance}
\label{tab:lgb}
\scriptsize
\begin{tabular}{l*{4}{c}}

\toprule
\\[-2ex] 
&{Precision}&{Recall}&{F1-Score}&{Accuracy}\\
\\[-2ex] 
\midrule
\\[-2ex] 
\textbf{Validation} \\
\cmidrule{1-1}
\\[-2ex]        
\textbf{LightGBM} & 0.869 & 0.872 & 0.870 & 0.821 \\
\\[-2ex]
\textbf{Uniform Baseline} & 0.689 &  1.0 & 0.816 & 0.689 \\
\\[-2ex]
\textbf{Random Baseline} & 0.690 & 0.568 & 0.623 & 0.526 
\\
\\[-2ex] 
\midrule
\\[-2ex] 
\textbf{Testing} \\
\cmidrule{1-1}
\\[-2ex]
\textbf{LightGBM} & 0.847 & 0.831 & 0.838 & 0.810 \\
\\[-2ex]
\textbf{Uniform Baseline} & 0.593 & 1.0 & 0.745 & 0.593 \\
\\[-2ex]
\textbf{Random Baseline} & 0.617 & 0.612 & 0.615 & 0.544 \\

\botrule
\end{tabular}
\begin{tablenotes}[flushleft]

\item Notes: Baseline use the same features as the best performing LightGBM Model. 

\end{tablenotes}
\end{table}

\begin{table}[h]
\centering 
\caption{Ablation study on GPT generated features}
\label{tab:lgb_ablation}
\scriptsize
\renewcommand\theadfont{}
\begin{tabular}{l*{4}{c}}

\toprule
\\[-2ex] 
&{Precision}&{Recall}&{F1-Score}&{Accuracy}\\
\\[-2ex] 
\midrule
\\[-2ex] 
\textbf{Testing} \\
\cmidrule{1-1}
\\[-2ex]
\textbf{Non-textual features} & 0.593   &  1.0 & 0.745  &  0.593 \\
\\[-2ex]
\textbf{Non-textual features + lexicon-based text features} & 0.717 & 0.931 & 0.810 & 0.741 \\
\\[-2ex]
\textbf{Non-textual features + GPT-generated features} & 0.781  & 0.779 & 0.780  & \textbf{0.739}\\
\\[-2ex]

\textbf{All features} & 0.847 & 0.831 & 0.838 & \textbf{0.810} \\
\\[-2ex]
\botrule
\end{tabular}


\end{table}

\clearpage
\newpage

\subsection{Regression Analysis}
\label{si-sec:logit}
To better understand the relationships between the key textual features (including lexicon-based text features and ChatGPT-generated features) and fundraising outcome, we run logistic regressions with Stata and present the results in Table \ref{tab:regression}. 

Model (1) only includes the campaign configuration features. The results suggest both fundraising goal amount ($\beta=0.03$, $p<.01$) and the organizer's gender (male) ($\beta=0.16$, $p<.01$) are positively correlated with fundraising success, which is different from prior findings showing insignificant positive relationship \cite{yazdani2024mis} or significant negative relationship \cite{peng2022speaking}. 

In Model (2), we additionally control for the features that describe local demographics and level of pandemic shock. Due to the high correlation among the ACS variables, we deploy a Principal Component Analysis to extract five principal components of local demographics, which together account for 68.5\% variances of all ACS features. Therefore, instead of including all the ACS variables, we include these five principal components of local demographics in Model (2). The results show that small businesses are more likely to receive financial support if they are located in cities with larger populations ($\beta=0.01$, $p<.01$), cities with lower average employment rates ($\beta =-0.02$, $p<.01$), and cities with less young residents ($\beta=-0.02$, $p<.01$). Additionally, we also observe that GoFundMe donors tend to prioritize small businesses in the areas experiencing relatively more severe pandemic impact (as shown by the percentages of a state's new COVID cases in relation to the total U.S. count, $\beta=0.03$, $p<.01$). The total number of new cases in the past week exhibits significant negative relationship with fundraising success ($\beta=-0.01$, $p<.01$), yet for which we would expect limited interpretation since this observed pattern may interplay with the timely trend that there are increasing number of COVID cases during the observation window. 


We additionally include textual features extracted from campaign descriptions in Model (3). These features are selected according to the literature and the feature importance scores of the best-performing LightGBM model. The results lead to interesting findings discussed in Section \ref{sec:discussion}.

\newgeometry{top=1.5cm}
\begin{table}[]
\centering
\caption{ Logistic regression on fundraising outcome.  }

\label{tab:regression}
\scriptsize
\begin{tabular}{l*{3}{c}}

\toprule

&\parbox[t][0.01cm][c]{0.05cm}{(1)}&\parbox[t][0.01cm][c]{0.2cm}{(2)}&\parbox[t][0.01cm][c]{0.2cm}{(3)}\\

& {\rule[0.5ex]{2cm}{0.4pt}}&{\rule[0.5ex]{1.4cm}{0.4pt}}&{\rule[0.5ex]{2cm}{0.4pt}}\\
                    
&{GFM Semantics}&{Context}&{Textual Features}\\
\\[-2ex] 
\midrule
\\[-2ex]  
GFM - goal amount     &        0.02\sym{***}&        0.01\sym{*}  &        0.00         \\
                    &      (0.00)         &      (0.00)         &      (0.00)         \\
GFM - male organizer  &        0.16\sym{***}&        0.16\sym{***}&        0.05\sym{***}\\
                    &      (0.01)         &      (0.01)         &      (0.01)         \\
GFM - has beneficiary     &        0.00         &       -0.02         &        0.11\sym{***}\\
                    &      (0.01)         &      (0.01)         &      (0.01)         \\

ACS population      &                     &        0.02\sym{***}&        0.01\sym{***}\\
                    &                     &      (0.00)         &      (0.00)         \\
ACS wealth      &                     &        0.00         &        0.00         \\
                    &                     &      (0.00)         &      (0.00)         \\
ACS employment      &                     &       -0.02\sym{***}&       -0.01\sym{***}\\
                    &                     &      (0.00)         &      (0.00)         \\
ACS youth           &                     &       -0.02\sym{***}&       -0.01\sym{***}\\
                    &                     &      (0.00)         &      (0.00)         \\
ACS racial diversity&                     &        0.00         &        0.01\sym{**} \\
                    &                     &      (0.00)         &      (0.00)         \\
\# of new COVID cases in the past week&                     &       -0.01\sym{**} &       -0.01\sym{***}\\
                    &                     &      (0.00)         &      (0.00)         \\
\# of new COVID cases / \# of new in U.S.&                     &        0.03\sym{***}&        0.03\sym{***}\\
                    &                     &      (0.01)         &      (0.01)         \\
1st person plural (we)&                     &                     &        0.01\sym{**} \\
                    &                     &                     &      (0.00)         \\
2st person (you)    &                     &                     &        0.02\sym{***}\\
                    &                     &                     &      (0.00)         \\
Word count          &                     &                     &        0.08\sym{***}\\
                    &                     &                     &      (0.01)         \\
Word per sentence   &                     &                     &       -0.00         \\
                    &                     &                     &      (0.00)         \\
Syllable per word   &                     &                     &       -0.01\sym{***}\\
                    &                     &                     &      (0.00)         \\
Concreteness level of words&                     &                     &        0.02\sym{***}\\
                    &                     &                     &      (0.00)         \\
Small business specified&                     &                     &        0.37\sym{***}\\
                    &                     &                     &      (0.01)         \\
Employees mentioned &                     &                     &        0.05\sym{***}\\
                    &                     &                     &      (0.01)         \\
Rent of business mentioned&                     &                     &        0.04\sym{***}\\
                    &                     &                     &      (0.01)         \\
Businesses longer than 2 years&                     &                     &        0.05\sym{***}\\
                    &                     &                     &      (0.01)         \\
New business        &                     &                     &       -0.05\sym{***}\\
                    &                     &                     &      (0.01)         \\
Match grant mentioned&                     &                     &        0.05\sym{***}\\
                    &                     &                     &      (0.02)         \\
Gratitude expressed &                     &                     &        0.06\sym{***}\\
                    &                     &                     &      (0.01)         \\
Urgency explained   &                     &                     &        0.09\sym{***}\\
                    &                     &                     &      (0.01)         \\
Spam words mentioned&                     &                     &       -0.04\sym{***}\\
                    &                     &                     &      (0.01)         \\
Dominance level of words&                     &                     &        0.03\sym{***}\\
                    &                     &                     &      (0.00)         \\
Social comparison $(better \ than \ peers)$&                     &                     &       -0.06\sym{***}\\
                    &                     &                     &      (0.02)         \\
Self comparison $(worse \ than \ self)$&                     &                     &        0.02\sym{*}  \\
                    &                     &                     &      (0.01)         \\
Extrinsic incentive        &                     &                     &        0.05\sym{***}\\
                    &                     &                     &      (0.02)         \\

\midrule 
McFaddnen's $\mathbf{R}^2$  & 0.02 & 0.05 & 0.29 \\[1mm]
\\[-2ex]         
\# of Descriptions &  11,274 &  11,274 &   11,274   \\

\botrule
\end{tabular}
\begin{tablenotes}[flushleft]

\item Notes: Average marginal effect with \textit{delta-method} SE in parentheses. Model (1) is used to control GoFundMe campaign configurations. Model (2) is employed to control contextual features, which include five principal components derived from a principal component analysis of local demographics and two features about pandemic shock. Model (3) includes both lexicon-based and ChatGPT generated textual features as discussed in Section \ref{sec:results} Findings.   \sym{*}   \(p<0.1\), \sym{**} \(p<0.05\), \sym{***} \(p<0.01\)


\end{tablenotes}

\end{table}
\restoregeometry

\newpage

\setlength\tabcolsep{3pt}
\newgeometry{top=1.5cm,left=2.5cm}
\begin{table}[h]
\centering 
\vspace{-8mm}
\caption{A brief summary of literature and our findings about lexicon-based and ChatGPT-generated features}
\label{tab:feature_literature_finding}
\vskip -8pt
\footnotesize
\begin{tabular}{l*{5}{l}}
\toprule
\\[-2ex] 
\textbf{Feature Group}&\textbf{\quad\quad\quad Example Features}&\textbf{\makecell[c]{Findings in \\ Charitable Giving Literature}}&\textbf{\makecell[c]{Findings \\ (LightGBM) }}&\textbf{\makecell[c]{Findings\\ (Logistic Regression)}}\\
\\[-2ex] 
\midrule
\\[-2ex] 
Textual & Word Count (LIWC) & Peng et al. (2022) - significant positive correlation & 2.69\%  & significant positive correlation \\
complexity & Word Per Sentence (LIWC) & Du et al. (2021): included in prediction model & 0.52\% & insignificant correlation \\
& Syllable Per Word (Textstat) & - & 0.28\% & significant positive correlation \\
& Reading Grade (Textstat) & Du et al. (2021): included in prediction model & 0.43\% & - \\
\\[-2ex] 
\midrule
\\[-2ex]
Linguistic  & Authentic (LIWC) & - & 0.08\% & -\\
style & Analytical thinking (LIWC) & Yazdani et al. (2024): insignificant positive correlation & 0.34\% & - \\
& Clout (LIWC) & \makecell[l]{Zhang et al. (2021): High clout (i.e., higher confident\\
but lower humble)  is negatively related to crowdfunding \\
fundraising success.} &  0.56\% & - \\
& Concreteness level of words & Du et al. (2021): included in prediction model & 0.06\% & significant positive correlation \\
& Dominance level of words & - & 1.62\% & significant positive correlation\\
& Spam words & - & 0.05\% & significant negative correlation \\
\\[-2ex] 
\midrule
\\[-2ex]
Pronouns, & 1st Person Plural-We (LIWC) & Zhu (2022): significant positive correlation & 0.38\% & significant positive correlation \\
punctuations, & 2st Person-You (LIWC) & Zhu (2022): significant positive correlation  & 
2.71\% & significant positive correlation \\
function & Numbers (LIWC) & Du et al. (2021): included in prediction model & 0.24\% & - \\
\\[-2ex] 
\midrule
\\[-2ex]
Cognition & Cognition (LIWC) & Mitra and Gilbert (2014): significant positive correlation & 0.2\% & - \\
& Causation (LIWC) & Peng et al. (2022): insignificant negative correlation & 0.17\% & - \\
& Discrepancy (LIWC) & Peng et al. (2022): significant negative correlation & 0.33\% & - \\
& Tentative (LIWC) & Peng et al. (2022): insignificant negative correlation & 0.18\% & \\
& Certitude (LIWC) & Peng et al. (2022): significant negative correlation & 0.15\% & \\
\\[-2ex] 
\midrule
\\[-2ex]
Emotion,   & Affect (LIWC) & Mitra and Gilbert (2014): significant positive correlation & 0.07\% & -  \\
sentiment, & Positive emotion (LIWC) & Yazdani et al. (2024) - significant positive correlation &  0.21\% & -\\
perception & Anxiety (LIWC) & \makecell[l]{Zhao et al. (2021) \& Peng et al. (2022) - significant\\ positive correlation} & 0.12\% & -  \\
& Sentiment Polarity (TextBlob) & -  &  0.27\% & - \\
&  Perception (LIWC) & Du et al. (2021): included in prediction model & 0.09\% & - \\
& Feeling (LIWC) & Peng et al. (2022) - significant positive correlation & 0.15\% & - \\
& Self Comparison (GPT) & - & 0.02\% & significant positive correlation \\
\\[-2ex] 
\midrule
\\[-2ex]
Social  & Prosocial behavior (LIWC) & - & 0.65\% & -\\
processes & Family (LIWC) & Peng et al. (2022) - no significant positive correlation & 0.07\% & - \\
& Friends (LIWC) & Peng et al. (2022) - no significant positive correlation & 1.65\% & - \\
& Social Comparison (GPT) & \makecell[l]{Frey and Meier (2004): experimental evidence on the \\ positive effect of donor social comparison on donation} & 0.02\% & \makecell[l]{significant negative correlation \\ of fundraiser peer comparison \\
on donation}\\
& Gratitude expressed (GPT) & \makecell[l]{Grant and Gino (2010): experimental evidence on the \\
positive effect of gratitude expression in university \\ alumni fundraising} & 0.98\% & significant positive correlation \\
\\[-2ex] 
\midrule
\\[-2ex]
Culture, & Culture (LIWC) & - & 0.11\% & -   \\
lifestyle& Politics (LIWC) & - & 0.15\% & - \\
& Lifestyle (LIWC) & - & 0.35\% & - \\
 \\[-2ex] 
\midrule
\\[-2ex]
Need,  & Health (LIWC) & Peng et al. (2022) - no significant positive correlation & 0.25\% & - \\
drives & Food (LIWC) & - & 2.02\% & - \\
& Acquire (LIWC) & - & 0.26\% & - \\
 \\[-2ex] 
\midrule
\\[-2ex]
Financial & Money (LIWC) & Du et al. (2021): included in prediction model & 1.04\% & -  \\
aspects & \makecell[l]{Match grant acknowledgement\\ (GPT)} & 
\makecell[l]{Chen et al. (2005) \& Karlan and List (2007): 
experimental \\ evidence on the positive effect of matching mechanism} & 0.26\% & significant positive correlation \\
& \makecell[l]{Budget for employee support\\ (GPT)} & - & 1.55\% & significant positive correlation  \\
& Budget for rent (GPT) & \makecell[l]{Chao (2017): experimental evidence of negative \\(crowdout) effect;  Alpizar et al. (2008): experimental \\ evidence of significant small positive effect; } & 0.28\% & significant positive correlation\\
& \makecell[l]{Providing gift cards or other \\ extrinsic incentive (GPT)} & Falk (2007): experimental evidence of positive effect & 0.03\% & - \\
 \\[-2ex] 
\midrule
\\[-2ex]
Time  & Past focus (LIWC) & Peng et al. (2022) - no significant negative correlation & 0.1\% & - \\
orientation & Present focus (LIWC) & Peng et al. (2022) - significant negative correlation & 0.22\% & - \\
 & Future focus (LIWC) & \makecell[l]{Peng et al. (2022) - no significant negative correlation; \\ Zhao et al. 2021 - significant positive correlation}& 0.17\% & - \\
 & Urgency explained (GPT) & Yazdani et al. (2024) - significant positive correlation & 0.04\% & significant positive correlation\\
  \\[-2ex] 
\midrule
\\[-2ex]
 Business  & New business (GPT) & - & 0.1\% & significant negative correlation\\
properties & Businesses Longer than 2 yrs (GPT) & - & 0.05\% & significant positive correlation
 \\
 & Small business specified (GPT) & -  & 36\% & significant positive correlation\\

\\[-2ex]
\botrule
\end{tabular}
\begin{tablenotes}[flushleft]

\item Notes: Zhang et al. (2021) included 92 LIWC-2015 features, which demonstrated predictive power. ``Findings (LightGBM)'' refers to the \\corresponding feature importance (in percentage) in our best-performing LightGBM model. We refer the audience to Table \ref{tab:regression} for more details \\ of the coefficients and significance levels associated with the ``Findings (Logistic Regression)''.
\end{tablenotes}

\end{table}
\restoregeometry
\setlength\tabcolsep{6pt}

\newpage

\subsection{Simulation Analysis}
\label{si:simulation}
\subsubsection{Textual feature selection.} To verify the effectiveness of our findings, we incorporate three factors highlighted from our primary results in a simulation analysis: explicit disclosure of matching grants, expression of gratitude, and explanation of urgency. We select these three factors because they allow us to modify campaign descriptions without compromising the authenticity, thereby validating the effectiveness of actionable insights for launching fundraising campaigns. 

\subsubsection{Simulation sample.} Next, we select a random sample of campaigns whose original descriptions deviate from the optimal setup across these three dimensions. According to Section \ref{sec:results}, an ideal fundraising campaign should explicitly disclose the information about grant matches, express gratitude to donors for their support, and explanation of why it's urgent to get financial support. Therefore, we randomly sample 500 campaigns from the pool of campaigns that fail to meet all these criteria (i.e., lacking explicit disclosure of matching grants, expression of gratitude, and urgency explanation) to conduct counterfactual simulation.

\subsubsection{Counterfactual analysis.} 
\label{si:simulation_rewrite}
Next, following the standard practices in counterfactual analysis \cite{ye2020predicting}, for each campaign in our simulation sample, we hypothetically ``correct'' sentence fragments in the campaign descriptions that relate to expressions of gratitude, grant matching, and urgency, while keeping all other wording and the campaign configuration unchanged. Specifically, we use "gpt-4-1106-preview" via ChatGPT API \cite{OpenAI} to (1) incorporate sincere expressions of gratitude towards potential donors, (2) highlight the \$500 matching policy provided by GoFundMe, and (3) emphasize the urgency of securing sufficient funds. 


To ensure that ChatGPT adds textual content without compromising the authenticity of the information, we utilize Chain-of-Thought prompting \cite{wei2022chain} to facilitate a series of intermediate reasoning steps. Specifically, we instruct ChatGPT to incrementally augment the original content and articulate the reasoning behind each modification. This approach ensures that the GPT-augmented version not only retains text in the original campaign description but also add the three desired factors in line with the original campaign description. See Table \ref{tab:prompt_simulation} for a detailed prompt example.

The expected fundraising outcomes of these hypothetical campaigns are simulated by predicting the probability of securing any funds using our best-performing LightGBM model. We then measure the effectiveness of description edits by comparing the success probabilities between campaigns with GPT-augmented descriptions and those with their original descriptions. The results are visualized in Figure \ref{fig:simulation}.

\begin{table}[h]
\centering 
\caption{Prompt for ChatGPT-augmented campaign descriptions }
\label{tab:prompt_simulation}
\begin{tabular}{{l}}

\toprule
\\[-2ex] 
\makecell[l]{
\textbf{PROMPT:}
\\
\\[-2ex]
You will be provided with a text delimited by triple quotes. The text comes from a crowdfunding campaign description. \\
It's trying to raise money for a business. You have these following tasks, please output the result in JSON format:\\
\\[-2ex]
\textbf{Task 1} : Given text: \textit{TEXT}, you need to do the following modifications to the original text and output the modified \\ version. 
Important note: For this task, the first part of your text will be exactly the same as the original text and then \\
you add sentences after the original text according to the description of each task.\\
\\[-2ex]
Firstly, you need to add two to three sentences to the original text to express sincere gratitude towards potential backers,\\
thanks for their help and kindness. Use a wealth of statements to express this gratitude. Make the sentence concrete \\
according to the original text. \\
\\[-2ex]
Secondly, Point out that if the author can raise \$500, GoFundMe's Small Business Relief Initiative will match \$500 \\
for the business, and this will be a huge help for the business. Use different sentences to express the meaning. Finally, \\
add two to three sentences to the original text to describe that the need of funds for the business is very urgent. Use a \\ wealth of statement's to express the urgency. Make sure the sentence is concrete 
according to the original text. Output \\
the modified sentences after these three steps of modification into the field  [correct\_three]. \\
\\[-2ex]
\textbf{Task 2} : Given text: \textit{TEXT}, Add two to three sentences to the original text to express sincere gratitude towards potential \\
backers, thanks for their help and kindness. Use a wealth of statements to express this gratitude. Make the sentence\\ concrete
according to the original text. The final result should contains the original text as well as the newly 
created \\
sentences. \textit{Important note}: For this task, the first part of your text will be exactly the same as the original text and \\
then you add sentences after the original 
text according to the description of each task. Output the result to \\
field [add\_gratitude].\\
\\[-2ex]
\textbf{Task 3}: Given text: \textit{TEXT}, tell me whether the text express gratitude to potential backers, and find which sentence does \\
so. Remove this sentence from the original text, and output the result to field [minus\_gratitude]. \\
\\[-2ex]
The returned json object should have the following four fields: correct\_three, add\_gratitude, minus\_gratitude. The value \\
of each field should be a string of sentence.
}\\
\\[-2ex]

\botrule
\end{tabular}
\end{table}

\clearpage
\newpage

\subsubsection{Simulation Analysis Robustness Check}
To rule out the alternative hypothesis that the observed effect of ChatGPT augmentation arises merely from an increase in text length, we conduct a robustness check using linear regression models as follows:

\begin{equation}
    y = \beta_1 \times \text{ChatGPT augmentation} + \beta_2 \times \text{text length} + \epsilon
    \label{eq:robust}
\end{equation}

\noindent where $y$ represents the predicted funding probability derived from the best-performing LightGBM model, and ChatGPT augmentation is a binary variable with the GPT-augmented campaigns coded as 1 and the original campaigns coded as 0. Text length is denoted as the word count of the campaign description.  As shown in Table \ref{tab:robust_check_three}, the results suggest that the effect of ChatGPT augmentation is statistically significant controlling for the text length ($\beta_1 = 0.09$, $p<.01$).


\begin{table}[!h]
\centering 
\caption{Robustness Check for ChatGPT augmentation that adds gratitude expression, matching grant acknowledgement, and urgency explanation to the original description.  }
\label{tab:robust_check_three}
\begin{tabular}{l*{3}{c}}

\toprule
& \multicolumn{3}{c}{Dependent Variable: Predicted Funded Likelihood}\\
\\[-2ex] 
\cmidrule{2-4}
&\parbox[t][0.01cm][c]{0.05cm}{(1)}&\parbox[t][0.01cm][c]{0.2cm}{(2)}&\parbox[t][0.01cm][c]{0.2cm}{(3)}\\

& {\rule[0.5ex]{2cm}{0.4pt}}&{\rule[0.5ex]{3cm}{0.4pt}}&{\rule[0.5ex]{3.4cm}{0.4pt}}\\

&{All}&{Funded Campaigns}&{Unfunded Campaigns}\\
\\[-2ex] 
\midrule
\\[-1ex]         

GPT Augmentation             &        0.12\sym{***}&        0.10\sym{***}&        0.13\sym{***}\\
                    &      (0.02)         &      (0.03)         &      (0.01)         \\
Word Count                  &        0.00         &       -0.07\sym{***}&        0.03\sym{***}\\
                    &      (0.01)         &      (0.02)         &      (0.01)         \\

\midrule

\makecell[l]{\# of Campaign Descriptions \\ (including both original and\\ GPT-augmented versions)} &  1,000 &  322 &   678  \\

\botrule
\end{tabular}
\begin{tablenotes}[flushleft]

\item Notes: SEs in parentheses. \sym{*}   \(p<0.1\), \sym{**} \(p<0.05\), \sym{***} \(p<0.01\)

\end{tablenotes}

\end{table}

\subsubsection{Social Inequity and ChatGPT Augmentation}

While crowdfunding platforms present alternative financing avenues for all small businesses, these avenues may not equally benefit all fundraisers \cite{kenworthy2020cross}. Figure \ref{fig:equity} visualizes demographic disparities in fundraising results. Small businesses from cities with a higher proportion of non-adults under 18 years, more African American citizens, and more residents in poverty are less likely to receive financial support (Figure \ref{fig:equity}a,c,e). We also find that percentage of foreign-born citizens (Figure \ref{fig:equity}b) and population density (Figure \ref{fig:equity}d) are positively correlated with the percentage of funded campaigns. In addition, our results suggest gender disparities where campaigns led by female organizers are less likely to be funded on GoFundMe.

\begin{figure}[H]
\centering
\includegraphics[width=\textwidth]{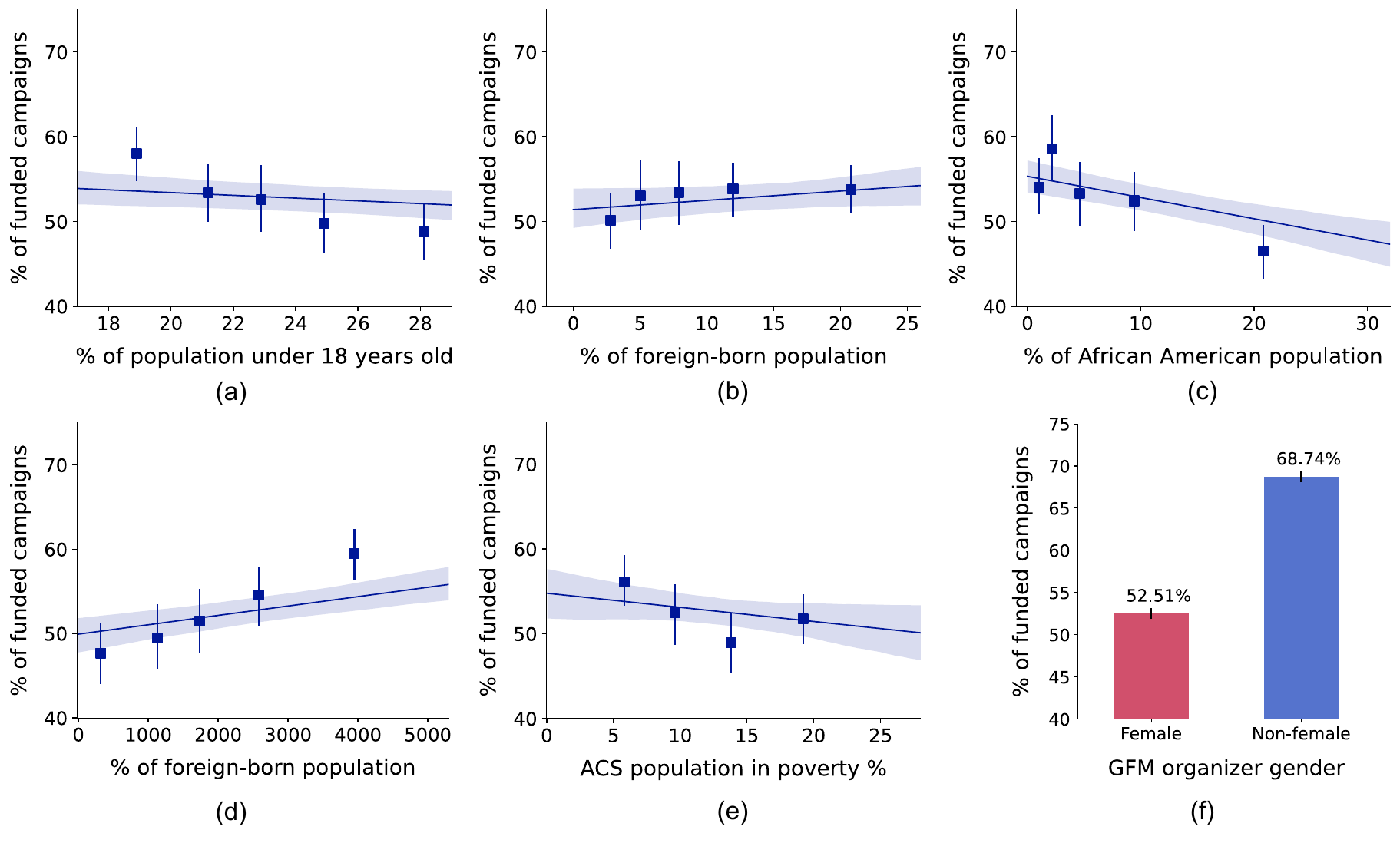}
\caption{Demographic disparities in crowdfunding outcome. Subplots (a) to (e) depicts the relationships between city demographics and the percentage of funded campaigns in the city. Specifically, (a) shows that campaigns from regions with a higher percentage of individuals under 18 years old are less funded. (b) indicates that small businesses in areas with a greater proportion of foreign-born citizens tend to achieve higher funded percentages. (c) reveals that areas with more African American residents see lower funding success for small businesses. (d) demonstrates that businesses in densely populated areas are more likely to receive funding. (e) suggests that lower funding rates are observed in areas with more population in poverty. (f) describes the percentage of funded campaigns for male and female campaign organizers (at the campaign level) and uncovers that campaigns organized by female are less likely to be funded, highlighting a gender disparity.}
\label{fig:equity}
\end{figure}

To explore the potential contribution of ChatGPT augmentation on crowdfunding equity, we conduct a counterfactual analysis and examine the heterogeneous effects of ChatGPT augmentation on enhancing the likelihood of receiving funding across regions with different levels of high education, as well as between male and female campaign organizers.

Specifically, we randomly select 1,000 campaigns from all campaigns that did not include expressions of gratitude in their original descriptions. We then apply ChatGPT to integrate gratitude expression while keeping all other narratives unchanged. We choose to alter gratitude expression since it is universally applicable to all campaigns and helps avoid introducing selection biases. 

Comparing the predicted funded probability before and after ChatGPT augmentation, our results from linear regression analyses (Table \ref{tab:social_inequity} Model (1)) show that small businesses from cities of lower education levels exhibits a higher increase in the likelihood to receive financial support: 1\% of increase in adult residents (older than 25 years) holding a Bachelor's or higher degree may translate to 0.34\% lower funded probability increase with ChatGPT augmentation ($p<.01$).  Adding campaign configuration features into Table \ref{tab:social_inequity} Model (2), we show that AI can help reduce the gap of crowdfunding success between male and female organizers: female organizers on average enjoy 17.65\% more benefit from ChatGPT augmentation ($p<.01$).

Such findings contribute to the growing discussion on the structural fundraising disadvantages faced by underrepresented groups (e.g., \citenum{coleman2009comparison,bapna2021gender,gafni2021gender,greenberg2015leaning}) and potential mechanisms to close the gap (e.g., \citenum{kanze2018we}). The proposed generative-AI approach highlights ground-breaking accessible opportunity to mitigate the educational and gender disparities.



\begin{table}[!h]
\centering 
\caption{Heterogeneous treatment effects of ChatGPT augmentation: Linear regressions}
\label{tab:social_inequity}
\begin{tabular}{l*{2}{c}}

\toprule
& \multicolumn{2}{c}{Dependent Variable: $\Delta$ of Predicted Funded Likelihood}\\
\\[-2ex] 
\cmidrule{2-3}
&\parbox[t][0.01cm][c]{0.2cm}{(1)}&\parbox[t][0.01cm][c]{0.2cm}{(2)}\\

& {\rule[0.5ex]{2.4cm}{0.4pt}}&{\rule[0.5ex]{3.8cm}{0.35pt}}\\

&{Education} & {With Campaign Configurations}\\
\\[-2ex] 
\midrule
\\[-1ex]         

Percentage of persons aged 25 years and older&       -0.34\sym{**} &       -0.37\sym{**} \\
with a bachelor's degree or higher                    &      (0.14)         &      (0.14)         \\
\\[-2ex] 
GFM - goal amount      &                     &        0.41\sym{***}\\
(in 1000 dollars)                    &                     &      (0.12)         \\
\\[-2ex] 
GFM - male organizer  &                     &       -17.65\sym{**} \\
                    &                     &      (4.98)         \\
\\[-2ex] 
GFM - has beneficiary     &                     &      -3.31 \sym{***}\\
                    &                     &       (4.89)          \\

\midrule

\makecell[l]{N} &  1,000 & 1000  \\

\botrule
\end{tabular}
\begin{tablenotes}[flushleft]

\item Notes: SEs in parentheses. \sym{*}   \(p<0.1\), \sym{**} \(p<0.05\), \sym{***} \(p<0.01\)

\end{tablenotes}

\end{table}

\newpage

\subsection{Online Experiments}
\label{si:experiment}
We conduct an online experiment to examine the effectiveness of the insights discovered from our predictive analysis. Below, we discuss the details of the experiment implementation. 

\subsubsection{Experimental conditions}
We design and implement three conditions in our experiment: (1) original campaigns, serving as the control group, which use the original campaign descriptions; (2) GPT-augmented campaigns, our treatment group, where campaign descriptions are revised using ChatGPT to incorporate additional information based on insights from predictive analysis, namely, expressions of gratitude, matching grant acknowledgement, and urgency explanations; and (3) GPT-extended campaigns, acting as a placebo condition, where ChatGPT paraphrases and extends the original introductions without adding new information. The third condition addresses concerns regarding the potential influence of text length differences on participants' preferences, and ensuring that participants are not biased by the GPT-augmented campaigns simply because they are longer than the original campaigns.  The prompt to generate GPT-extended campaign introduction is presented in Table \ref{tab:prompt_extension}.
\begin{table}[h]
\centering 
\caption{Prompt to generate GPT-extended campaign descriptions}
\label{tab:prompt_extension}
\begin{tabular}{{l}}

\toprule
\\[-2ex] 
\makecell[l]{
\textbf{PROMPT:}
\\
\\[-2ex]
You will be provided with text deliminated by triple quotes. The text comes from a crowdfunding campaign \\ description.
It's trying to raise money for its business. Provided text: \textit{TEXT}. Can you add about \textit{LENGTH} \\ emotion-neutral words
in the following text by paraphrasing the sentences in the original text and make sure  \\ the new text satisfies the following
conditions:\\
\\[-2ex]
(1)  the first sentence should be the same as the first sentence from the original text, \\
(2) the last sentence should be the same as the last sentence from the original text,\\
(3) please generate the new text with four paragraphs, make sure the beginning of each paragraph using exact \\ same sentence extracted from the original text, \\
(4) while the total length of the text increases, keep everything else the same, such as you should not adding any \\facts 
or information, don't overstate the business scale or service quality and not changing emotional valence, \\and don't give examples that are not included in the original text. \\
\\[-2ex]
Output result of modified text.}\\
\\[-2ex]

\botrule
\end{tabular}

\end{table}

\subsubsection{Campaign selection and campaign description generation}
\label{sec:exp_sample_selection}
To select campaigns for the experiment, we employ stratified sampling to randomly select 16 campaigns from the 500 used in our simulation analysis. Specifically, we first divide these 500 campaigns into two groups based on whether they had successfully secured funding. Next, to ensure it is feasible to extend the original campaigns to match the length of the GPT-augmented versions without introducing new information, we select only those campaigns with an original length of more than 180 words. \footnote{This approach provides a more conservative estimate of the effect size since longer original campaigns have less room for improvement through ChatGPT augmentation.} Finally, to improve the representativeness of our sample, we create four strata within each of the funded and unfunded groups, based on the difference in predicted fundraising probability before and after ChatGPT augmentation. From each stratum of funded and unfunded campaigns, we randomly select 2 campaigns, resulting in a total of 16 campaigns for our sample.


\subsubsection{Randomization}
\label{sec:randomization}
Each participant in our experiment is presented with two random pairs of campaign introductions. Specifically, for each participant, we first randomly draw one campaign that was successfully funded and one that did not secure funding. For each selected campaign, we then randomly choose one pair of campaign-description variations from the three types: original, GPT-augmented, and GPT-extended. This means, for each campaign-description evaluation, participants compare two out of these three variations. The order in which the funded and unfunded campaign pairs are presented to participants is random. Within each pair, the two variations of the same campaign are displayed side by side, also in a random order. 

\subsubsection{Dependent variables}
We consider two variables to examine donation preference. Each participant is asked to indicate (1) their own donation preference
towards either Campaign 1 or Campaign 2 in each pair, and (2) their prediction of most people's preference. The latter variable has been claimed to provide a more accurate estimate of general donation behavior \cite{prelec2004bayesian}).

\subsubsection{Attention Check}

To ensure attentive participation and comprehension of our experimental materials, we incorporate two types of attention checks: a general attention check at the beginning of the survey, and a campaign-specific attention check following the evaluation of each pair of campaign introductions.

\textit{General attention check}. We follow the guidelines in the literature \cite{oppenheimer2009instructional} and include an instructional attention check task at the beginning of our experiment (see Figure \ref{fig:attentioncheck1} for details). This page contains a title ``Introduction to GoFundMe'', an instructional paragraph, and a multiple-choice question. The instruction paragraph starts with a detailed introduction to GoFundMe and concludes with an instruction for participants to click the title to continue, rather than answering the multiple-choice question and clicking ``Next page'' to proceed. Participants who adhere the instruction demonstrate sufficient attention and pass the attention check. Positioned as the survey's very first task, only those who pass this attention check are allowed to proceed to the evaluation of the crowdfunding campaigns. 

\textit{Attention check specific to campaigns}. To ensure that participants understand the campaign introduction, campaign-specific attention checks ask about the business category of the campaign, after participants review the campaign. See Figure \ref{fig:attentioncheck2} for an example.
\begin{figure}[H]
\centering
\includegraphics[width=0.7\textwidth]{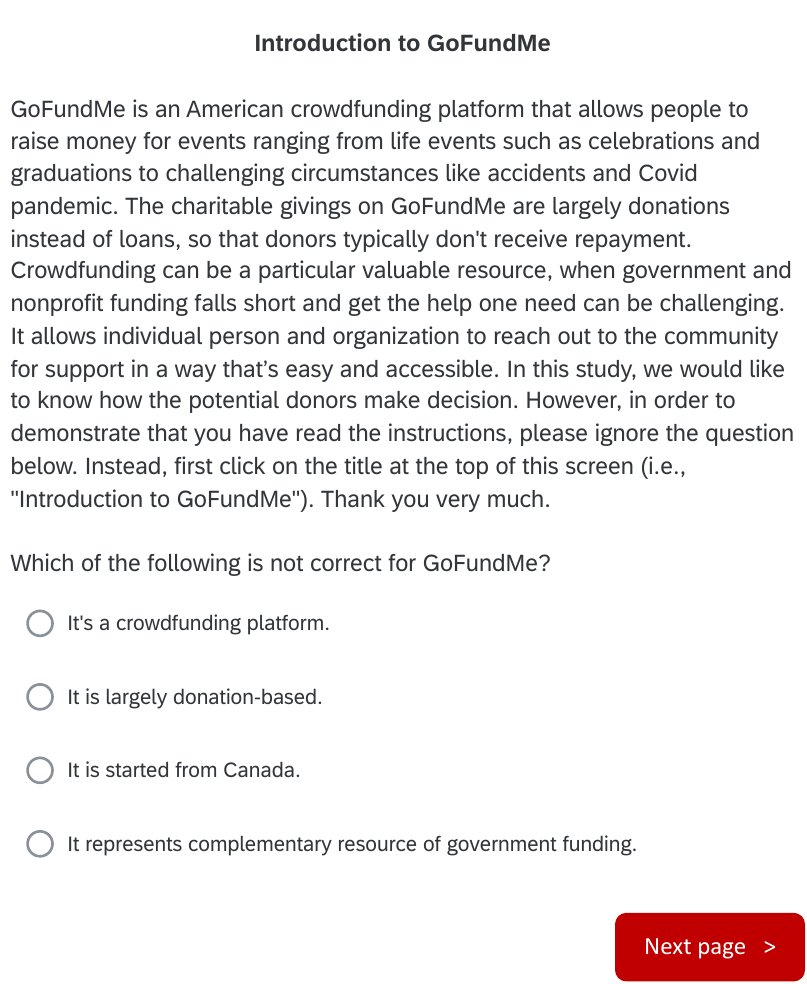}
\caption{Attention check 1}
\label{fig:attentioncheck1}
\end{figure}

\begin{figure}[H]
\centering
\includegraphics[width=0.6\textwidth]{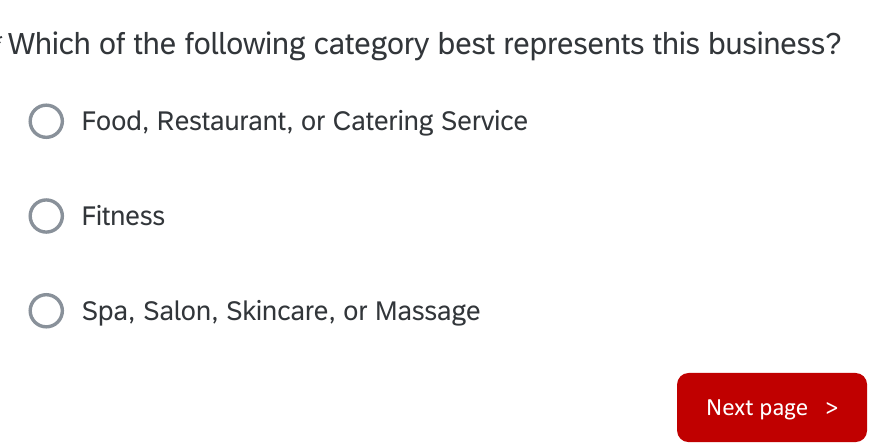}
\caption{Attention check 2}
\label{fig:attentioncheck2}
\end{figure}

\subsubsection{Experimental procedure}

In summary, if a participant opts to take the survey, they conduct a one-time online survey with the following steps:

\begin{itemize}
\item Step 1: General attention check. Only participants who pass the attention check may proceed.
\item Step 2: Evaluation of the first pair of campaigns. The first pair of campaign descriptions is presented side by side. Tables \ref{tab:example_sub1} - \ref{tab:example_sub3} show examples of campaign pairs.
\item Step 3: Participants indicate their own donation preference towards either Campaign 1 or Campaign 2, estimate which campaign will attract more donations from the general public, and explain their reasoning (see Figure \ref{fig:survey_question} for an example). Following this, participants complete a campaign-specific attention check.  
\item Step 4: Evaluation of the second pair of campaigns. The second pair of campaign descriptions is presented side by side. 
\item Step 5: Participants indicate their own donation preference towards either Campaign 1 or Campaign 2, estimate which campaign will attract more donations from the general public, and explain their reasoning. Following this, participants complete a campaign-specific attention check. 
\item Step 6: Participants fill out a brief post-experiment questionnaire, including age, gender, and how many times they have donated in the past year. 

Participants are then thanked and debriefed, which concludes the study. 
\vskip-10pt
\begin{figure}[H]
\centering
\includegraphics[width=0.58\textwidth]{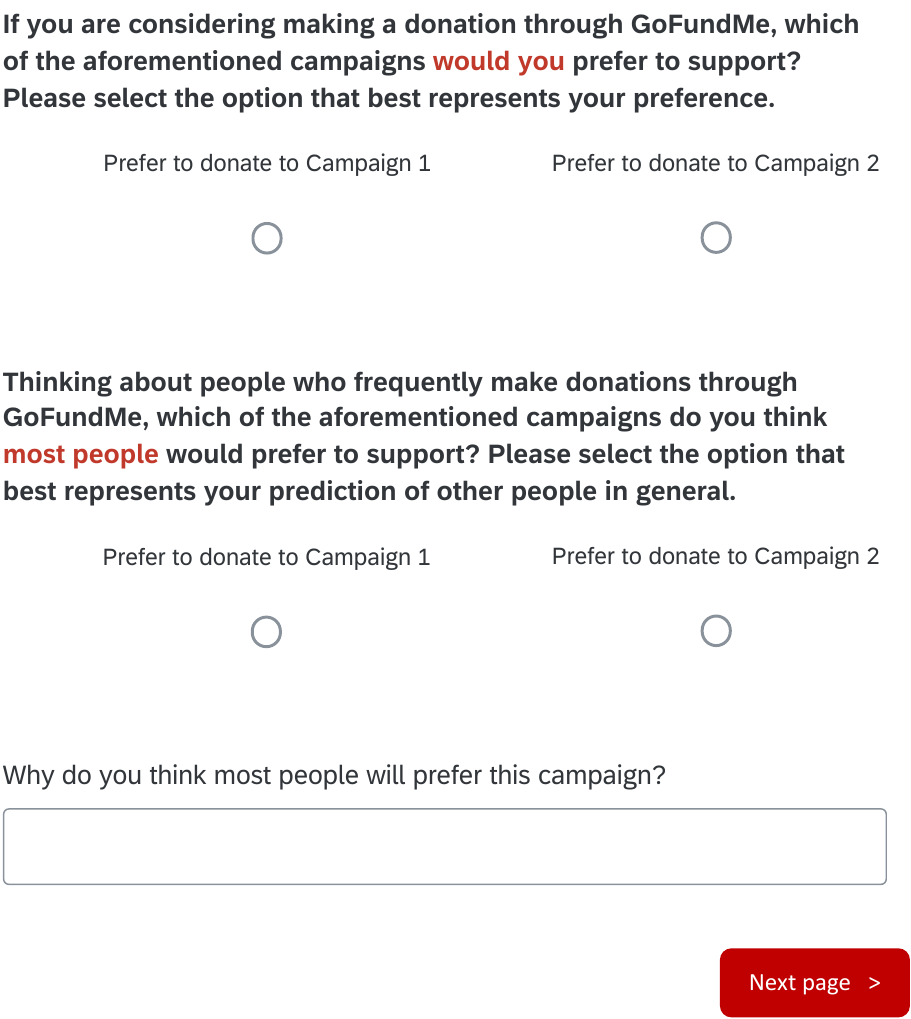}
\caption{Example of Step 3 in the experimental procedure.}
\label{fig:survey_question}
\end{figure}

\end{itemize}
\vskip-20pt

\newgeometry{left=2.5cm}
\begin{table}[h]
\centering 
\caption{An example of original (left) versus GPT-extended (right) campaign descriptions}
\label{tab:example_sub1}
\begin{tabular}{c|c}

\toprule
\\[-2ex]
\makecell[l]{\textbf{Campaign 1}} & \makecell[l]{\textbf{Campaign 2}}\\
\midrule
\\
\textbf{Support Silky Smooth Waxing Studio COVID-19 Relief}&\textbf{Support Silky Smooth Waxing Studio COVID-19 Relief}\\
\\[-2ex] 

\\[-2ex]         
\makecell[l]{\hspace{5mm} We specialize in Brazialian Waxing for women. \\
\hspace{5mm} 
The studio offers private and custom skin care and gentle\\
effective waxing in an atmosphere that is slow-paced, nurturing\\
and private. We offer body waxing treatment from head to toe \\
and everywhere in between. Our technicians are highly qualified;\\
trained in Europe and USA. \\
\hspace{5mm} 
We are meticulous with sanitary measures. Angel (Zdravka \\
Simova) is the owner and one of the licensed estheticians at \\
the Studio. She got her first Esthetician diploma in Europe \\
and later graduated with honors from Hanover Park College of \\
Beauty Culture to obtain her Illinois State License. She also\\
took her post-graduate degree in Skin Care Education at The \\
Rayner Institute for Career Advancement (which focuses \\
specifically on Dermatologic skin care) to become Certified \\
Clinical Estehtician. \\
\hspace{5mm} 
She continues to further her education in this growing \\
industry to keep up with the latest advancements in skin care. \\
She has applied her knowledge working in the Midwest for one \\
of the nation's top 200 salons, where she performed top-notch,\\
quality waxing services and skin care treatments. She has \\
exceptionally high standards for the delivery of client services,\\
and her performance levels are always on par with her true \\
capabilities. She has found that clients have high expectations, \\
and she is eager to meet and exceed them!\\
~\\~\\~\\~\\~\\~\\~\\~\\~\\~\\~\\~\\~\\~\\}& 
\makecell[l]{\hspace{5mm} We specialize in Brazilian Waxing for women. \\
\hspace{5mm} 
The studio offers private and custom skin care and gentle \\
effective waxing in an atmosphere that is slow-paced, nurturing \\
and private. Our establishment is dedicated to providing an \\
environment that supports tranquility and privacy, ensuring that \\
each client receives personalized skin care tailored to their \\
individual needs. We pride ourselves on our gentle and effective \\
waxing techniques that cater to a diverse clientele seeking \\
comprehensive body waxing treatments that range from the top of \\
the head to the tips of the toes, and all areas in between. Our
team \\
of technicians boasts an impressive level of qualification, 
having \\
received extensive training across both European and
American \\
methodologies. \\
\hspace{5mm} 
Our technicians are highly qualified; trained in Europe and \\
USA. They bring a wealth of knowledge and expertise to the studio, \\
ensuring that every treatment is performed with the utmost precision \\
and care. The studio's commitment to hygiene and cleanliness is \\
unwavering, with meticulous attention to sanitary measures that \\
adhere to the highest standards in the industry. Angel (Zdravka \\
Simova), 
the proprietor and a licensed esthetician at the Studio, is \\
a testament to the caliber of professionals we have on board. Her \\
educational journey began in Europe, where she earned her initial \\
Esthetician diploma, and continued in the United States, where \\
she graduated with honors from Hanover Park College of Beauty \\
Culture to secure her Illinois State License. \\
\hspace{5mm} 
She also took her post-graduate degree in Skin Care Education \\
at The Rayner Institute for Career Advancement (which focuses \\
specifically on Dermatologic skin care) to become Certified Clinical \\
Esthetician. Angel's pursuit of excellence in the field of skin care is \\
evident in her ongoing commitment to education, as she consistently \\
seeks out new advancements in the industry to enhance her knowledge \\
and skills. Her experience is extensive, having honed her craft in the \\
Midwest at one of the nation's top 200 salons, where she delivered \\
high-quality waxing services and skin care treatments. Angel's \\ standardsfor client service are exceptionally high, ensuring that \\
every experience at the studio aligns with her professional ethos \\
and reflects her true capabilities. She has found that clients have \\
high expectations, and she is eager to meet and exceed them!
}\\
\\
\botrule
\end{tabular}
\end{table}
\restoregeometry

\newgeometry{left=3cm}
\begin{table}[h]
\centering 
\caption{An example of GPT-extended (left) versus GPT-augmented (right) campaign \\descriptions}
\label{tab:example_sub2}
\begin{tabular}{c|c}

\toprule
\\[-2ex]
\makecell[l]{\textbf{Campaign 1}} & \makecell[l]{\textbf{Campaign 2}}\\
\midrule
\\
\textbf{Support Silky Smooth Waxing Studio COVID-19 Relief}&\textbf{Support Silky Smooth Waxing Studio COVID-19 Relief}\\
\\[-2ex] 
\\[-2ex]         
\makecell[l]{\hspace{5mm} We specialize in Brazilian Waxing for women. \\
\hspace{5mm} 
The studio offers private and custom skin care and gentle \\
effective waxing in an atmosphere that is slow-paced, nurturing \\
and private. Our establishment is dedicated to providing an \\
environment that supports tranquility and privacy, ensuring that \\
each client receives personalized skin care tailored to their \\
individual needs. We pride ourselves on our gentle and effective \\
waxing techniques that cater to a diverse clientele seeking \\
comprehensive body waxing treatments that range from the top of \\
the head to the tips of the toes, and all areas in between. Our
team \\
of technicians boasts an impressive level of qualification, 
having \\
received extensive training across both European and
American \\
methodologies. \\
\hspace{5mm} 
Our technicians are highly qualified; trained in Europe and \\
USA. They bring a wealth of knowledge and expertise to the studio, \\
ensuring that every treatment is performed with the utmost precision \\
and care. The studio's commitment to hygiene and cleanliness is \\
unwavering, with meticulous attention to sanitary measures that \\
adhere to the highest standards in the industry. Angel (Zdravka \\
Simova), 
the proprietor and a licensed esthetician at the Studio, is \\
a testament to the caliber of professionals we have on board. Her \\
educational journey began in Europe, where she earned her initial \\
Esthetician diploma, and continued in the United States, where \\
she graduated with honors from Hanover Park College of Beauty \\
Culture to secure her Illinois State License. \\
\hspace{5mm} 
She also took her post-graduate degree in Skin Care Education \\
at The Rayner Institute for Career Advancement (which focuses \\
specifically on Dermatologic skin care) to become Certified Clinical \\
Esthetician. Angel's pursuit of excellence in the field of skin care is \\
evident in her ongoing commitment to education, as she consistently \\
seeks out new advancements in the industry to enhance her knowledge \\
and skills. Her experience is extensive, having honed her craft in the \\
Midwest at one of the nation's top 200 salons, where she delivered \\
high-quality waxing services and skin care treatments. Angel's \\ standardsfor client service are exceptionally high, ensuring that \\
every experience at the studio aligns with her professional ethos \\
and reflects her true capabilities. She has found that clients have \\
high expectations, and she is eager to meet and exceed them!
}& 
\makecell[l]{\hspace{5mm} We specialize in Brazilian Waxing for women. \\
\hspace{5mm} 
The studio offers private and custom skin care and gentle \\
effective waxing in an atmosphere that is slow-paced, nurturing \\
and private. We offer body waxing treatment from head to toe and \\
everywhere in between. Our technicians are highly qualified; \\
trained in Europe and USA. \\
\hspace{5mm} 
We are meticulous with sanitary measures. Angel (Zdravka \\
Simova) is the owner and one of the licensed estheticians at the \\
Studio. She got her first Esthetician diploma in Europe and later \\
graduated with honors from Hanover Park College of Beauty Culture \\
to obtain her Illinois State License. She also took her post-graduate\\
degree in Skin Care Education at The Rayner Institute for Career \\
Advancement (which focuses specifically on Dermatologic skin care) \\
to become Certified Clinical Esthetician. \\
\hspace{5mm} 
She continues to further her education in this growing \\
industry to keep up with the latest advancements in skin care. She \\
has applied her knowledge working in the Midwest for one of the \\
nation's top 200 salons, where she performed top-notch, quality \\
waxing services and skin care treatments.  She has exceptionally \\
high standards for the delivery of client services, and her \\
performance levels are always on par with her true capabilities. \\
She has found that clients have high expectations, and she is eager \\
to meet and exceed them! \\
\hspace{5mm} 
We are deeply grateful for the support and generosity of \\
our potential backers. Your help is not just a contribution, it's \\
a statement of belief in the quality and care we provide. Every \\
pledge brings us closer to our goal and reinforces our commitment \\
to excellence in skin care and waxing services.\\
\hspace{5mm} 
If we are able to raise \$500, we have the incredible opportunity \\
to have that amount matched by GoFundMe's Small Business Relief \\
Initiative. This matching grant would be a significant boost for \\
our business, effectively doubling the impact of your donations. \\
\hspace{5mm}
The need for these funds is pressing. We are at a critical juncture \\
where every contribution can make the difference in sustaining \\
our high-quality services and the livelihood of our dedicated staff. \\
Your timely support is crucial and will be instrumental in helping \\
us navigate through these challenging times.\\~\\
}\\
\\

\botrule
\end{tabular}

\end{table}
\restoregeometry

\newgeometry{left=3cm}
\begin{table}[h]
\centering 
\caption{An example of GPT-augmented (left) versus original (right) campaign descriptions}
\label{tab:example_sub3}
\begin{tabular}{c|c}

\toprule
\\[-2ex]
\makecell[l]{\textbf{Campaign 1}} & \makecell[l]{\textbf{Campaign 2}}\\
\midrule
\\
\textbf{Support Silky Smooth Waxing Studio COVID-19 Relief}&\textbf{Support Silky Smooth Waxing Studio COVID-19 Relief}\\
\\[-2ex] 
\\[-2ex]       
\makecell[l]{\hspace{5mm} We specialize in Brazilian Waxing for women. \\
\hspace{5mm} 
The studio offers private and custom skin care and gentle \\
effective waxing in an atmosphere that is slow-paced, nurturing \\
and private. We offer body waxing treatment from head to toe and \\
everywhere in between. Our technicians are highly qualified; \\
trained in Europe and USA. \\
\hspace{5mm} 
We are meticulous with sanitary measures. Angel (Zdravka \\
Simova) is the owner and one of the licensed estheticians at the \\
Studio. She got her first Esthetician diploma in Europe and later \\
graduated with honors from Hanover Park College of Beauty Culture \\
to obtain her Illinois State License. She also took her post-graduate\\
degree in Skin Care Education at The Rayner Institute for Career \\
Advancement (which focuses specifically on Dermatologic skin care) \\
to become Certified Clinical Esthetician. \\
\hspace{5mm} 
She continues to further her education in this growing \\
industry to keep up with the latest advancements in skin care. She \\
has applied her knowledge working in the Midwest for one of the \\
nation's top 200 salons, where she performed top-notch, quality \\
waxing services and skin care treatments.  She has exceptionally \\
high standards for the delivery of client services, and her \\
performance levels are always on par with her true capabilities. \\
She has found that clients have high expectations, and she is eager \\
to meet and exceed them! \\
\hspace{5mm} 
We are deeply grateful for the support and generosity of \\
our potential backers. Your help is not just a contribution, it's \\
a statement of belief in the quality and care we provide. Every \\
pledge brings us closer to our goal and reinforces our commitment \\
to excellence in skin care and waxing services.\\
\hspace{5mm} 
If we are able to raise \$500, we have the incredible opportunity \\
to have that amount matched by GoFundMe's Small Business Relief \\
Initiative. This matching grant would be a significant boost for \\
our business, effectively doubling the impact of your donations. \\
\hspace{5mm}
The need for these funds is pressing. We are at a critical juncture \\
where every contribution can make the difference in sustaining \\
our high-quality services and the livelihood of our dedicated staff. \\
Your timely support is crucial and will be instrumental in helping \\
us navigate through these challenging times.
}&
\makecell[l]{\hspace{5mm} We specialize in Brazialian Waxing for women. \\
\hspace{5mm} 
The studio offers private and custom skin care and gentle\\
effective waxing in an atmosphere that is slow-paced, nurturing\\
and private. We offer body waxing treatment from head to toe \\
and everywhere in between. Our technicians are highly qualified;\\
trained in Europe and USA. \\
\hspace{5mm} 
We are meticulous with sanitary measures. Angel (Zdravka \\
Simova) is the owner and one of the licensed estheticians at \\
the Studio. She got her first Esthetician diploma in Europe \\
and later graduated with honors from Hanover Park College of \\
Beauty Culture to obtain her Illinois State License. She also\\
took her post-graduate degree in Skin Care Education at The \\
Rayner Institute for Career Advancement (which focuses \\
specifically on Dermatologic skin care) to become Certified \\
Clinical Estehtician. \\
\hspace{5mm} 
She continues to further her education in this growing \\
industry to keep up with the latest advancements in skin care. \\
She has applied her knowledge working in the Midwest for one \\
of the nation's top 200 salons, where she performed top-notch,\\
quality waxing services and skin care treatments. She has \\
exceptionally high standards for the delivery of client services,\\
and her performance levels are always on par with her true \\
capabilities. She has found that clients have high expectations, \\
and she is eager to meet and exceed them!\\
~\\~\\~\\~\\~\\~\\~\\~\\~\\~\\~\\~\\~\\}\\

\\

\botrule
\end{tabular}

\end{table}
\restoregeometry

\subsubsection{Recruited participants}
We recruit participants from an online survey platform, Prolific. To ensure the similarity between our recruited sample and GoFundMe donors, we follow the literature \cite{athey2022smiles} and only include participants with high socio-economic status (i.e., those with a household annual income exceeding $80,000$). Furthermore, we limit our study to participants located in the United States to ensure they understand the scale of donation volume.

From a total of 307 participants, we remove those who fail the instructional attention check and filter out the campaign-pair records where the participants cannot correctly recall the business category. As a result, our final sample consists of 263 participants, with an average age of 42.59 years and 44.1\% female. In total, these participants evaluate 507 valid pairs of campaigns. Table \ref{tab:survey_demographics} shows the summary of the demographics of our participants. For randomization check, we conduct pairwise Kolmogorov-Smirnov tests across three combinations of campaign comparisons (GPT-augmented versus GPT-extended, GPT-augmented versus original, and GPT-extended versus original), and find that none of the pairwise tests is significant at the conventional level ($p > 0.10$).

\begin{table}[h]
\centering 
\caption{Summary of demographics}

\label{tab:survey_demographics}
\begin{tabular}{lcc}
\toprule
& Mean & Std \\ 
\midrule
Gender (male = 0, female = 1) & 0.44 & 0.50 \\ 
Age & 42.59 & 13.48 \\
Donated in last year (never = 0, 1-3 times = 1,  $\textgreater$ times = 2) & 1.24 & 0.67  \\
Donated amount in last year (\$0-100 = 0, \$101-500 = 1, $\textgreater$ \$500 = 2) & 0.62 & 0.77 \\
\midrule
\# of Participants & 263 \\
\botrule
\end{tabular}




\end{table}

\subsubsection{Experiment hypotheses testing}

We develop three hypotheses to examine the effect of GPT augmentation on fundraising success. All these hypotheses have been pre-registered \cite{ye2023ai}. 
\begin{hyp}
Donors will be more likely to provide financial support to campaigns with GPT-augmented introductions than those with GPT-extended introductions.
\end{hyp}

\begin{hyp}
Donors will be more likely to provide financial support to campaigns with GPT-augmented introductions than those with original introductions.
\end{hyp}

\begin{hyp}
Donors will be more likely to provide financial support to campaigns with GPT-extended introductions than those with original introductions.
\end{hyp}

The utility of subject $i$ choosing option $j$ can be represented as follows:
\begin{equation*}
u_{ij} = \beta_0 + \beta_1 * \text{GPT Augmentation} + \beta_2 * \text{GPT Extension} + \epsilon
\end{equation*}
where GPT Augmentation and GPT Extension are binary variables that respectively represent the GPT-augmented and GPT-extended conditions. Hypothesis 1 (or hypothesis 2) is considered as supported if $\beta_1>0$ (or $\beta_2>0$). Hypothesis 3 indicates $\beta_1>\beta_2$.

Following previous literature on discrete choice models \cite{athey2022smiles}, we test our hypotheses by estimating the average treatment effect using logistic regression maximizing the conditional likelihood. 

As shown in Table \ref{tab:survey_logistic}, participants are 34\% more likely to donate to the GPT-augmented campaigns than the original versions ($p<.01$), supporting hypothesis 1. GPT-extended campaigns outperform the original campaigns by 7\% ($p<.01$), confirming hypothesis 2. In addition, the significance difference between $\beta_1$ and $\beta_2$ validate hypothesis 3. These results are further corroborated  by participants' predictions about the preferences of the general public. 

Moreover, to further ensure that our participants represent donor profiles, we conduct a robustness check by restricting our analysis to those who have donated at least once in the past year. The robustness check confirms our hypotheses, showing statistically significant effects ($p<.01$) with $\beta_1 = 0.35$ and $\beta_2 = 0.09$ for participants' own donation preference, and $\beta_1 = 0.37$ and $\beta_2 = 0.08$ for participants' predictions of others' donation preferences.

\begin{table}[h]
\centering 
\caption{Conditional Logistic Regression for Online Experiment}

\label{tab:survey_logistic}
\begin{tabular}{l*{4}{c}}

\toprule

&\parbox[t][0.01cm][c]{0.05cm}{(1)}&\parbox[t][0.01cm][c]{0.05cm}{(2)}&\parbox[t][0.01cm][c]{0.05cm}{(3)}&\parbox[t][0.01cm][c]{0.05cm}{(3)}\\

&{\rule[0.5ex]{2cm}{0.4pt}}& {\rule[0.5ex]{2cm}{0.4pt}}&{\rule[0.5ex]{2.4cm}{0.4pt}}&{\rule[0.5ex]{2.4cm}{0.4pt}}\\

&{Self preference}&{Public prediction}&{ \makecell[c]{Self preference\\ (donated last year)}}&{\makecell[c]{Public prediction \\ (donated last year)}}\\
\\[-2ex] 
\midrule
\\[-1ex]         

GPT Augmentation    & 0.34\sym{***}  & 0.36\sym{***} & 0.35\sym{***} &  0.37\sym{***}  \\
   ($\beta_1$)                   &  (0.02)       &  (0.02)       &  (0.03)        & (0.03)        \\
GPT Extension    & 0.07\sym{***} & 0.07\sym{***} &  0.09\sym{***} & 0.08\sym{***} \\
  ($\beta_2$)                    &  (0.03)       &  (0.03)       &  (0.03)        &  (0.03)       \\[1ex]

\midrule
\\[-2ex]    
 $H_0$  ($\beta_1 = \beta_2$)  & \(p<0.01\) & \(p<0.01\) &  \(p<0.01\)  & \(p<0.01\)
  \\[0.5ex]     
\midrule

\# of Descriptions &  1,014 &  1,014 &  862  &  862  \\

\botrule
\end{tabular}
\begin{tablenotes}[flushleft]

\item Notes: Average marginal effect with \textit{delta-method} SE in parentheses. \sym{*}   \(p<0.1\), \sym{**} \(p<0.05\), \sym{***} \(p<0.01\)

\end{tablenotes}
\end{table}

\end{appendices}

\end{document}